\documentclass[sigconf]{acmart}

\definecolor{darkgreen}{rgb}{0,0.5,0}
\definecolor{orange}{rgb}{1,0.5,0}
\definecolor{teal}{rgb}{0,0.5,0.5}
\definecolor{olive}{rgb}{0.6,0.6,0}

\usepackage{xspace}
\newcommand {\verbalFramework}{Verbal Structuring\xspace}
\newcommand{\systemname}[1][]{\textsc{Orality}#1\xspace}

\newif\ifusecolor
\usecolorfalse

\ifusecolor
    \newcommand{\cl}[1]{\textcolor{blue}{#1}}
    \newcommand{\wx}[1]{\textcolor{blue}{#1}}
    
\else
    \newcommand{\cl}[1]{#1}
    \newcommand{\wx}[1]{#1}
    
\fi

\newif\ifshowdeleted
\showdeletedfalse

\newcommand{\del}[1]{%
    \ifshowdeleted
        \textcolor{brown}{\sout{#1}}%
    \else
    \fi
}

\AtBeginDocument{%
  \providecommand\BibTeX{{%
    \normalfont B\kern-0.5em{\scshape i\kern-0.25em b}\kern-0.8em\TeX}}}

\usepackage{soul}
\usepackage{multirow}
\usepackage{subcaption}
\usepackage{caption}
\usepackage{bm}
\usepackage{fancybox}
\usepackage{graphicx}
\usepackage[ruled,vlined]{algorithm2e}

\copyrightyear{2026}
\acmYear{2026}
\setcopyright{cc}
\setcctype{by}
\acmConference[CHI '26]{Proceedings of the 2026 CHI Conference on Human Factors in Computing Systems}{April 13--17, 2026}{Barcelona, Spain}
\acmBooktitle{Proceedings of the 2026 CHI Conference on Human Factors in Computing Systems (CHI '26), April 13--17, 2026, Barcelona, Spain}
\acmDOI{10.1145/3772318.3791713}
\acmISBN{979-8-4007-2278-3/2026/04}

\begin{document}
\title{\texorpdfstring{\systemname}{Orality}: A Semantic Canvas for Externalizing and Clarifying Thoughts with Speech}

\author{Wengxi Li}
\authornote{Both authors contributed equally to this research.}
\orcid{0009-0003-1876-6303}
\affiliation{%
\department{School of Creative Media}
  \institution{City University of Hong Kong}
  \city{Hong Kong}
  \country{China}
}
\email{wengxili-c@my.cityu.edu.hk}

\author{Jingze Tian}
\authornotemark[1]
\orcid{0000-0001-8528-5276}
\affiliation{%
\department{School of Creative Media}
  \institution{City University of Hong Kong}
  \city{Hong Kong}
  \country{China}
}
\email{jingztian2-c@my.cityu.edu.hk}

\author{Can Liu}
\authornote{Corresponding Author}
\orcid{0000-0003-3267-3317}
\affiliation{%
\department{School of Creative Media}
  \institution{City University of Hong Kong}
  \city{Hong Kong}
  \country{China}
}
\email{canliu@cityu.edu.hk}






\renewcommand{\shortauthors}{Wengxi Li, Jingze Tian, Can Liu}

\begin{abstract}

People speak aloud to externalize thoughts as one way to help clarify and organize them. Although Speech-to-text can capture these thoughts, transcripts can be difficult to read and make sense due to disfluencies, repetitions and potential disorganization. To support thinking through verbalization, we introduce \systemname{}, which extracts key information from spoken content, performs semantic analysis through LLMs to form a node-link diagram in an interactive canvas. Instead of reading and working with transcripts, users could manipulate clusters of nodes and give verbal instructions to re-extract and organize the content in other ways. It also provides AI-generated inspirational questions and detection of logical conflicts. We conducted a lab study with twelve participants comparing \systemname against speech interaction with ChatGPT. We found that \systemname{} can better support users in clarifying and developing their thoughts. The findings also identified the affordances of both graphical and conversational thought clarification tools and derived design implications.

\end{abstract}

\ccsdesc[500]{Human-centered computing~Interactive systems and tools}
\ccsdesc[500]{Human-centered computing~Interaction techniques}

\keywords{Speech-to-text, Human-AI Interaction, Large Language Models, Thought clarification, Tools for thinking}



\begin{teaserfigure}
  \includegraphics[width=\textwidth]{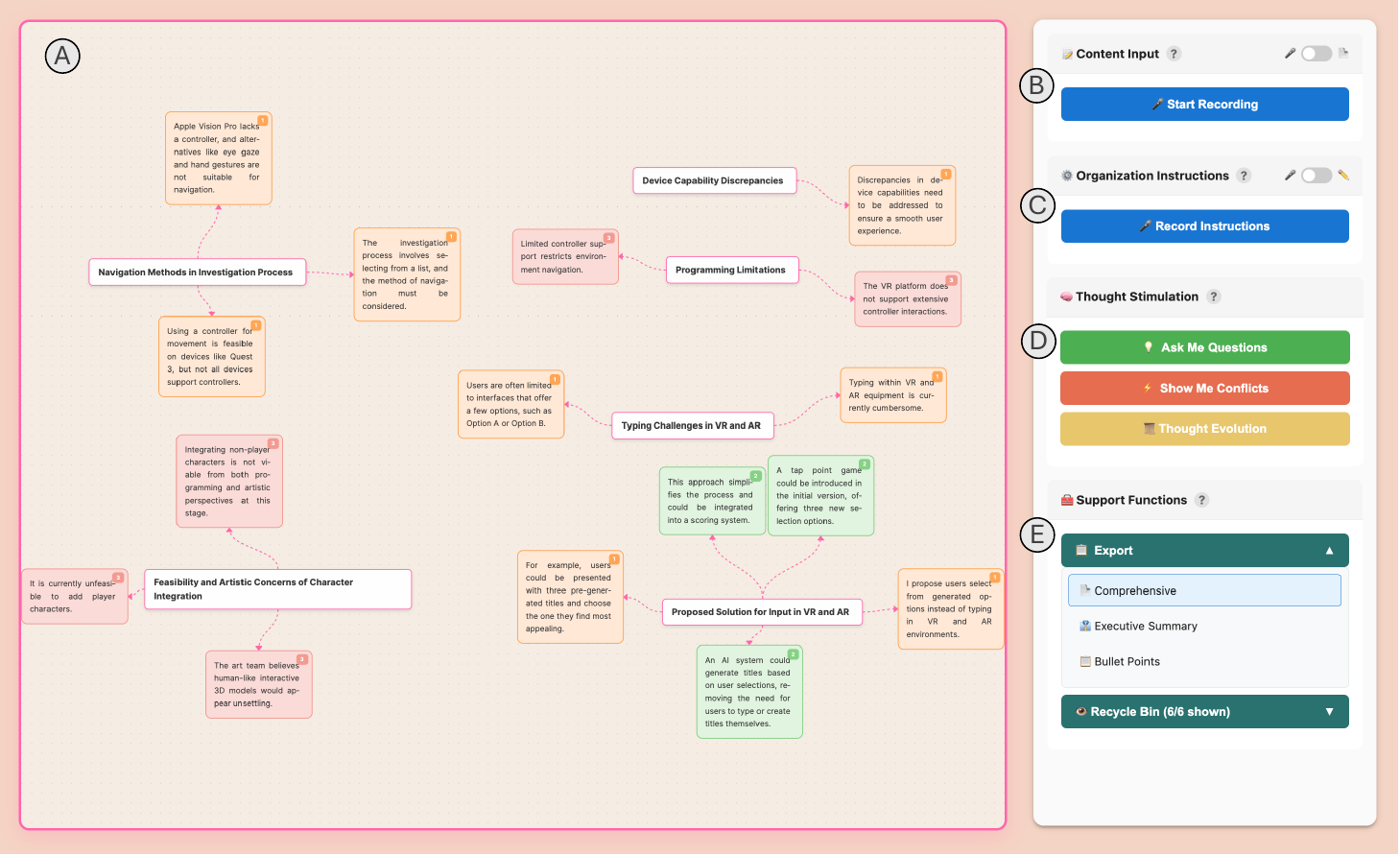}
  \caption{The main interface and features of \systemname with example content nodes and topics extracted from speech. Through the speech input widget (B), users can externalize their thoughts iteratively to the node-link semantic canvas (A). Users can adjust the structure and layout by verbally giving instructions (C). There are also thought stimulation functions (D) that can generate thought-guided question nodes by ``Ask Me Questions\del{Give Me Ideas},'' draw links between conflicting nodes by ``Show Me Conflicts,'' or navigate the thought evolution history by ``Thought Evolution.'' Users can also export memos for selected content (E).}
  \Description{A comprehensive interface view showing (a) the main thought mapping workspace with numerous interconnected text nodes and (b) the control panel featuring recording options, organization tools, thought stimulation buttons, and export functions, along with a series of small network diagrams illustrating the evolution of idea connections over time.}
  \label{fig:teaser}
\end{teaserfigure}


\maketitle
\section{Introduction} 
The process of developing complex ideas, from forming a research question to structuring a business strategy, is fundamentally tied to our ability to externalize thought. We rarely think in a vacuum; instead, we engage in a dynamic dialogue with our own ideas by offloading them onto external representations like sketches, notes, and diagrams~\cite{kirsh2010thinking}. This concept of extended cognition, where external artifacts become an integral part of our cognitive processes, highlights the importance of the tools we use to think~\cite{clark1998extended}. These tools have ranged from physical whiteboards to digital mind-mapping software, each offering a way to stabilize fleeting thoughts so they can be inspected, organized, and refined.

Among the various methods for externalizing thoughts, speech is a unique and powerful channel. It is a natural and rapid medium for expressing the stream of internal monologue, bypassing the mechanical and physical friction of typing or drawing~\cite{ruan2018comparing}. Psychological research has long shown that verbalizing one's thought process -- the ``think-aloud'' method -- improves problem-solving by helping individuals identify gaps in their logic and transform abstract concepts into concrete expressions~\cite{ericsson1980verbal,chi1994eliciting}. Spoken language can capture the characteristic structure of spontaneous thoughts, which often emerges in related ``clumps'' of ideas followed by discontinuous ``jumps'' to new topics~\cite{sripada2020structure}. However, speech as a potential medium for externalization also has some challenges. The output is inherently linear, disorganized, and unstructured, making it difficult to review, reorganize, or synthesize into a coherent whole~\cite{arons1997speechskimmer,chandrasegaran2019talktraces}. Although there are speech-to-text tools on most operational platforms today and recent transcription tools like otter.ai, Zoom, and notes apps on our phones can generate summaries, action items, and topic lists to provide overviews, the output still lacks either clarity or flexibility to be well-used as a thinking medium~\cite{asthana2025summaries}. 

Recent advancements in large language models (LLMs) present a new opportunity to address these challenges. People increasingly use generative AI not merely as an information retrieval tool, but as a partner for thinking and reflection~\cite{lev2024metacognitive,gerlich2025ai}. Users may discuss complex personal decisions or long-term goals with an AI, using the dialogue to examine problems from multiple perspectives and progressively clarify their own ambiguous thoughts~\cite{lin2024decision,Daryanto2025conversate}. When combined with speech input, these interactions can become more natural and conversational, lowering the barrier to externalization~\cite{lin2024rambler}. However, the design of current AI-based conversational interfaces limits their potential as effective tools for thought. Interfaces like voice-version ChatGPT are constrained by their linear, chat-based format. As a user speaks and the conversation grows, the sequential log of text becomes difficult to review and synthesize, and the AI can struggle to maintain context, often producing verbose outputs that obscure rather than clarify the user's core ideas~\cite{huang2024conversation}. 
This linear structure is not optimal to represent the web-like connections between thoughts. While AI-generated mindmaps have been explored as a solution, they still lack the flexibility and quality needed to capture the nuance of a developing thought process~\cite{homer2025comparative,schicchi2025closer}. Some works have explored the combination of using speech and canvas to help externalize and organize thoughts with flexibility. But such systems focus on supporting group brainstorming or real-time collaborative dialogue mapping in meetings, ~\cite{zhang2025ladica, chen2025meetmap}
rather than supporting individuals' reflective thinking. 

In this paper, we aim to design and develop a novel interface that supports the use of speech to externalize thoughts and help develop them.  
The primary goal is to aid 
the process of externalizing, structuring and stimulating one's unclear or complex ideas. To ground our design, we conducted a formative study to understand the difficulties users face when organizing their verbalized thoughts with a chatbot interface. We found that users struggled with the mismatch between their non-linear thinking and an outline-like linear structure, and they expressed a need for more salient and flexible ways to view and manipulate thought structure.


Based on these findings, we introduce \systemname, a novel system that integrates a speech-first workflow with an AI-organized semantic canvas to support the full cycle of thought clarification, which helps users transform vague initial ideas into concrete, well-structured concepts. We achieve this by supporting iterative refinement and spatial reorganization of distilled spoken content on a semantic canvas.
The design of \systemname is based on a four-layer conceptual framework that adapts the Pirolli and Card sensemaking model to the context of personal thought clarification~\cite{pirolli2005sensemaking}. We replace the information foraging loop with a \textit{Thought Externalization Layer} supported by 
speech input, and support a sensemaking loop as a \textit{Structuring and Schematizing Layer} supported by a dynamic, node-based data representation on a canvas. This is augmented by an \textit{Elaboration and Deepening Layer} that uses AI-powered features to help users identify gaps and conflicts, and a final \textit{Reflection and Presentation Layer} that aims to support metacognition and the synthesis of interim or final output. This integrated approach aims to transform a linear stream of spoken thought into a graphical, malleable, and reflective thinking workspace.

We conducted a lab user study with twelve participants to evaluate \systemname's features, usefulness, workflow and support for thinking processes, in comparison with a baseline -- a chat-based LLM interface with custom prompts for generating outlines and concept maps. 
The results of this study indicate that \systemname provides better support for thought clarification than the baseline. Our qualitative findings provide insights into how our speech-driven malleable canvas AI affords different thinking support in divergent and in-depth thinking, active sensemaking and thought development, from a chat-based AI. 
In summary, we make the following contributions:
\begin{itemize}
    \item The iterative design of \systemname, grounded by an adapted theoretical framework for self thought clarification and a formative study uncovering design opportunities 
    for supporting the clarification of thoughts through iterative verbalization.  
    \item The research artifact \systemname for supporting reflective sensemaking,  featuring an AI-assisted speech-to-text canvas that extracts gists from spoken thoughts into a manipulatable node-link diagram and asks thought-provoking questions in-place.  
    \item A within-subject lab study with twelve participants identifying the advantages of a multimodal canvas AI interface in supporting active and in-depth thinking, compared with a chat-based AI. 
\end{itemize}

\section{Related Work} ~\label{sec:related_work}

\subsection{Thought Externalization in Various Modalities}
Human thinking is an active process that often extends beyond the brain and into the environment. The theory of extended cognition suggests that we use external artifacts -- from simple pen and paper to digital devices -- as integral parts of our cognitive system~\cite{clark1998extended}. The initial step is thought externalization: transferring fleeting, internal ideas into a stable, external form, which is a mind-demanding task~\cite{sweller1988cognitive}. In the dual-process theory, there are two different cognitive processes: ones that are unconscious, rapid, automatic, and high capacity (called System 1), and those that are conscious, slow, and deliberative (called System 2)~\cite{kahneman2011thinking, evans2008dual}. The direct flow of early ideas aligns with fast, intuitive System 1 processing and is often based on abstract gist, but the act of recording them can impose a high cognitive load and demand a slow, analytical System 2 approach to create a verbatim representation~\cite{reyna1995fuzzy}. The choice of modality and tool is therefore critical to reducing this cognitive friction.

HCI research has explored various tools with a range of modalities to support externalization, such as writing, sketching, gesturing, and speaking. Externalizing thoughts by writing through a keyboard and text editor is the most common~\cite{flower1981writing, lee2024writing}. Digital note-taking applications, outliners, and mind-mapping tools like Mindalogue~\cite{zhang2024mindalogue} and VISAR~\cite{zhang2023visar} provide canvases for users to record thoughts through writing. Some other systems explored using immersive environments to spatialize ideas~\cite{xing2025immersed}. They demonstrate that spatial organization and direct manipulation are critical for helping users discover patterns, construct a mental model, and develop more ideas.

In addition to these visual and text-based systems, recent work has begun to explore speech as a medium to externalize thoughts. Foundational work in this area, such as Bolt's ``Put-That-There'' system, demonstrated how speech and gesture could be combined as a natural modality for controlling graphical interfaces~\cite{bolt1980putthatthere}. Rambler~\cite{lin2024rambler} supports writing long-form text via dictation and uses LLMs to help users organize and edit the content. LADICA~\cite{zhang2025ladica} and MeetMap~\cite{chen2025meetmap} extract information from the speech of group conversations into a canvas to facilitate their brainstorming activities. However, these tools tend to focus on assisting the production of certain output, e.g., articles or code, by either post-processing user-generated content or suggesting ideas from outside resources. There is a lack of tools and studies that focus on capturing, clarifying and structuring unclear thoughts, especially by using the speech input modality. 

Our work aims to support a more seamless and iterative transition between externalization and organization of thought,
from the rapid, associative externalization characteristic of System 1 to the deliberate, analytical organization of System 2, by creating an interface that seamlessly integrates both modes of thought. We believe a speech-based semantic canvas is a powerful support for this purpose. Our system leverages speech to create a node-link canvas that can show the semantics within the spoken content, which can assist users in externalizing, organizing, and iterating their thoughts from the very beginning of the creative process.

\subsection{AI-Assisted Canvas Interfaces for Text Sensemaking}
The process of clarifying one's thoughts is not a simple act of transcription but an active, constructive process of sensemaking~\cite{weick1995sensemaking}. Once ideas are externalized, a person can begin to organize, connect, and refine them into a coherent structure. HCI research has a rich history of building visualization tools that support this sensemaking loop. Some visual sensemaking tools allow users to spatially arrange notes to form thematic clusters and conceptual maps, e.g., TalkTraces for verbal content, VideoMap for videos, Sensecape and Graphologue for LLM-generated text~\cite{chandrasegaran2019talktraces,lin2024videomap,suh2023sensecape,jiang2023graphologue}. They have introduced capabilities to automatically find connections and provide a space for thought exploration, supporting the schema-building part of the sensemaking loop. Systems such as Luminate and Patchview can generate stories based on the sensemaking of multiple semantic dimensions~\cite{suh2024luminate,chung2024patchview}. Other systems like ConceptScope and ConceptEVA focus on helping users iteratively define and evolve their conceptual understanding of a topic~\cite{zhang2023concepteva,zhang2021conceptscope}. These tools provide powerful support for organizing rich information. 



While these visualization techniques are effective for analyzing external data, it remains an open area of exploration how they can be adapted to support the more personal and fluid process of thought sensemaking. This process is different because the information does not exist beforehand; it is generated and structured at the same moment. While traditional content creation tools, like text editors or outliners, inherently combine externalization and organization, they often lack the visual and semantic structuring capabilities of modern sensemaking tools. The opportunity lies in connecting the fluid, initial act of thinking aloud with the power of a dynamic, flexible visual canvas.

Our work aims to bridge this gap by creating an iterative loop between thought externalization and sensemaking. By using speech for fluid and efficient input, our system allows users to see their thoughts appear as manipulable external objects between iterations and supports highly flexible reorganization with verbal commands. This design aims to lower the barrier between articulating an idea and manipulating it, encouraging a more dynamic cycle of speaking, seeing, and structuring. This approach adapts AI-assisted visualization techniques to the unique challenge of helping a user make sense of their own emerging ideas.




\subsection{Supporting Thinking with Metacognitive Tools}
Effective tools for thought extend beyond capturing and organizing information; they should also support the underlying cognitive processes of thinking itself. A useful framework for understanding this support is metacognition, which refers to the capacity to monitor and control one's own cognitive processes~\cite{flavell1979metacognition}. Metacognition involves two main components: monitoring, which includes assessing one's understanding and progress (e.g., ``Is this argument coherent?''), and control, which involves directing cognitive activities to achieve a goal (e.g., ``I should explore this idea further before continuing'')~\cite{nelson1990metamemory}. Designing tools that explicitly support metacognition remains a central challenge in HCI.

The introduction of modern computational tools, and especially Generative AI (GenAI), has placed new and significant metacognitive demands on users. As Tankelevitch et al. argue, interacting with GenAI systems requires a high degree of metacognitive effort: users must continuously engage in planning, such as formulating effective prompts; monitoring, by evaluating the accuracy, relevance, and quality of the generated outputs; and control, by deciding when to trust the AI, when to revise its output, and how to integrate it into a larger workflow~\cite{lev2024metacognitive}. The HCI community is actively exploring how to understand, protect, and augment human cognition in this new era of AI-supported work~\cite{tankelevitch2025understanding}. One approach is to offer cognitive scaffolding, where the system provides structure and guidance to help users manage their process. For example, some systems use metacognitive prompts to encourage users to engage in more critical thinking when searching with GenAI~\cite{singh2025enhancing}. Other work has shown that scaffolding can be effective in learning environments for improving problem-solving and metacognitive skills~\cite{wang2023scaffolding}. Another approach is to increase system transparency, helping users build a more accurate mental model of the tool's capabilities and limitations, which in turn improves their ability to monitor and control their interactions~\cite{liao2023aitransparency}. These systems aim to make the user more aware of their own thinking process and to provide the means to direct it more effectively.

Our work contributes to this line of research by designing a system that lowers the metacognitive load at the early stage of thought formation. Some researchers propose that AI should act as an ``extraheric,'' a partner that helps foster higher-order thinking skills~\cite{yatani2024aiextraherics}. Our system aligns with this view by providing thought stimulation functions to support rather than dominate the thinking process. Besides, the automatic generation of a semantic canvas provides visual feedback on the structure and relationships within the externalized thoughts, directly supporting the monitoring process.
\section{Formative Study}

We conducted a formative study to uncover design challenges and opportunities for providing additional interaction and visual support in clarifying verbalized thoughts. The findings from this study informed the features implemented in our proposed prototype.

\subsection{Tasks, Participants and Procedure}

The study asked participants to experience the workflow of creating verbalized content using speech-to-text input and organizing it by prompting an LLM. Before the task, participants were instructed to choose an open-ended topic based on a recent event that required thought clarification (e.g., structuring a confusing project, reflecting deeply on an issue). This ensured that the content came from their recent memory and leveraged their real need for thought clarification.

Eight participants, five who self-reported as female and three as male, were recruited from a research institute. All participants have experience using speech-to-text tools and regularly use LLMs for daily tasks. Each of the participants was compensated with \$15 for their participation. Each study session took about one hour.

In the first half of the study, participants proposed a topic and were given at least 30 minutes to verbalize and organize their thoughts until they reached the most complete and optimal version. 
During the task, they used Google Docs’ speech-to-text function to input content and crafted prompts for an LLM product DeepSeek\footnote{https://www.deepseek.com/} to organize the transcripts as needed. 

In the second half, participants were interviewed to reflect on their experience, focusing on the need for additional support 
during verbalization and LLM-based organization. The entire session was recorded, transcribed, and analyzed using thematic analysis to identify design challenges and opportunities.

\subsection{Findings and Discussion}

This formative study explored how participants externalized and organized their thoughts for various self-proposed tasks, including project planning (P1, P2, P4), creative writing (P3, P5, P6), and problem-solving (P7, P8). Their verbalized content typically captured a free flow of participants' internal thinking, including fragmented ideas, half-formed utterances, and spontaneous sparks of thought, resembling an unfiltered think-aloud process. A representative example transcript was shown in Appendix \ref{Formative Example}. From our analysis of their process, we identified four key themes that revealed critical challenges and design opportunities.

\subsubsection{The Need for Non-linear Thought Representation.}
The most fundamental challenge reported was the inherent conflict between the non-linear nature of thought and the sequential nature of dictation. This linear stream of raw thought created a stream of text with an intertwined content structure that is hard to read, understand, and manage. Participants described their initial process as speaking ``\textit{whatever comes to mind}'' (P2), resulting in rambling content that was difficult to manage. P1 explicitly stated that their ``\textit{rambling content cannot be easily deleted or hierarchically categorized}.'' This led to a clear desire for a different representation. As P1 suggested, the ideal interface would allow them to see the LLM's feedback to ``\textit{understand the topic being discussed and any missing points},'' pointing to the need for topic modeling and an easy-to-grasp overview 
rather than a blob of transcript.

\subsubsection{The Need for Flexible, Hierarchical and User-Defined Structures.}
Participants consistently attempted to impose their own structure on their thoughts, either before or during thought verbalization. 
This happened when they started with a high-level planning of the content before diving into the specifics.
For example, P2 explicitly began by outlining the talk’s overall scaffold: ``\textit{what I am gonna say about the topic mainly includes: what to do, the research questions, what and how to design the experiment, the timeline, ...}'' before fleshing out the content details. Similarly, P5 first planned the narrative ``\textit{blueprint}''— setting, characters, and their roles— to ensure a coherent story before developing the actual content. These structures were highly task-dependent and user-driven. P1 organized an ongoing project using a three-part structure (past progress, current status, next steps), while P4 planned an event by ``\textit{filling in what to do along a timeline}.''
Furthermore, participants required support for hierarchical organization. P1's request for an ``overall arrangement'' and P2's use of a self-entered, multi-level outline demonstrated a need for tools that support both high-level and detailed views of their thought structure.

\subsubsection{The Need for Supporting an Iterative Thinking Process.} 
The study revealed that thought clarification is not a linear, one-shot process but a highly iterative one. P1 and P5 both noted that their thinking deepened through the act of ``\textit{retelling}'' or re-articulating their points, with P5 stating that self-reflection during the oral presentation allowed for ``\textit{deeper reflection and ideas on the topic}.'' This iterative process faces significant challenges with the chat-based interface. Participants struggled to merge new input with existing content without creating redundancies. For example, P1 described how they ``\textit{tried to explain some things in a coherent way, but the effect wasn't good (there was a lot of repetition), so I decided to organize them all together after finishing them}.'' Similarly, P5 found that the LLM's attempts at merging new paragraphs were ineffective, so they ``\textit{passively accepted the piecemeal organization}.'' This demonstrated a clear need for better support for iterative and controllable content evolution.

\subsubsection{The Need for Non-intrusive LLM Assistance and User Agency.}
While participants universally valued the LLM's ability to extract key points from their speech, such as ``\textit{it extracted points I mentioned in a passing comment and forgot}'' 
(P2), they also faced significant challenges with its assistance. A recurring theme was the trade-off between helpful automation and the loss of user control. Participants reported that the LLM could misinterpret their intent, leading to ``\textit{nonsense, repetitive content}'' (P1, P2), or over-interpret their ideas (P3, P4, P5). For creative tasks, P5 found the LLM's supplementary content to be ``\textit{too rigid}'' and that it could even lead them ``\textit{astray from their original train of thought}.'' P1 articulated the ideal role for the AI, stating they wished the LLM could provide feedback, indicate what was missing, and prompt for more detail. This suggests a potential need for intelligent support that provides valuable suggestions without overriding the user's control or creative direction.

\subsection{Design Goals}
Based on the findings from our formative study and support from relevant theories, we synthesized a set of four design goals to guide the development of our system. These goals are intended to directly address the key challenges and opportunities identified in the user's speech-driven thought clarification process.

\paragraph{\textbf{DG1: Support Semantic Inspection of Spoken Thought.}} To address the conflict between linear speech and non-linear thinking, we aim to automatically transform sequential speech into a 2D spatial representation. \cl{This is to support thinking with external representation ~\cite{kirsh2010thinking} by offloading the memory and computational load of holding and developing the thought structure in the mind alone ~\cite{scaife1996external}. By mapping verbal data to a visual space, we hope to support users in inspecting semantic relationships and the ``shape'' of their arguments visually. } 
\paragraph{\textbf{DG2: Provide a Flexible Representation for Verbalizing and Reorganizing Structure.}} Participants required dynamic structures that evolve with their thinking. \cl{Previous work also showed that speech-based text composition often externalizes a stream of consciousness, thus generating an evolving structure~\cite{lin2024rambler}.} Therefore we would like to enable users to define, apply, and iteratively restructure their organizational schema using flexible verbal commands, allowing the \wx{ 
structure to transit from vague to clear as users' internal mental model evolves.}
\paragraph{\textbf{DG3: Visualize the Iterative Changes of Thought Evolution.}} Since thinking deepens through revision, we aim to make this evolutionary process visible and navigable. \cl{Social science research has long demonstrated the importance of assisting professionals to ``Reflect-in-Action''~\cite{schon1983reflective} by allowing them to pause, reframe their perspective, and question their own thought process. To support Metacognitive Reflection~\cite{flavell1979metacognition}, the system should allow users to review and compare historical versions of their thought structure to evaluate their progress and refine their logic.}  
\paragraph{\textbf{DG4: Provide Intelligent Content Support While Maintaining User Agency.}} To address the trade-off between assistance and user intent, we aim to provide on-demand support that augments rather than dominates. \wx{Prior research in Mixed-Initiative Interaction~\cite{allen1999mixinitiative} suggests that systems should provide value (e.g., identifying gaps or conflicts) without taking away from the user's initiative. Our system aims to} scaffold thinking rather than taking control of the content or disrupting creative flow.

\section{Designing ORALITY -- an AI-assisted Thought Clarification Tool}
Drawing on the findings of the formative study and the design goals, we developed \systemname, an AI-enhanced semantic canvas that supports thought clarification by enabling users to iteratively externalize, refine, and reorganize their ideas through speech
(Figure~\ref{fig:teaser}). The remainder of this section is organized as follows: first we will introduce the theory behind the design in Section~\ref{sec:designrational}; then Section~\ref{sec:features} describes the key features of \systemname; and Section~\ref{sec:scenario} presents an example scenario demonstrating the use of \systemname for a research project; finally, Section~\ref{sec:implementation} provides information on the implementation details.

\begin{figure*}[t!]
  \centering
  \includegraphics[width= \textwidth]{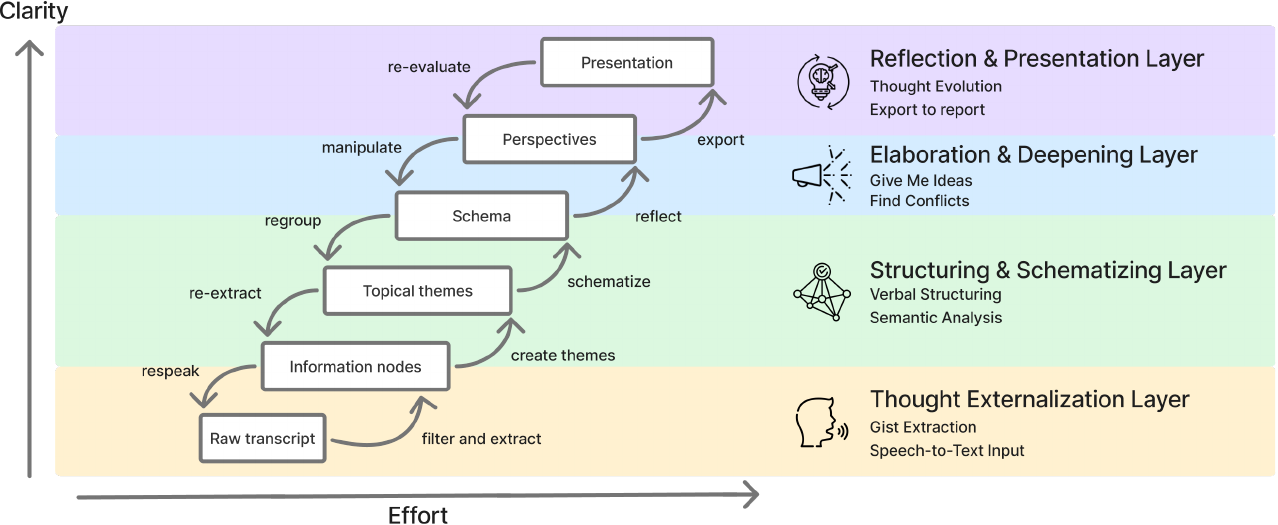}
  \caption{Our conceptual framework for self thought clarification process, which is adapted from the Sensemaking Model by Pirolli and Card~\cite{pirolli2005sensemaking}, involving stages corresponding to the four layers of system support designed in \systemname. Each layer lists the associated system features.}
  \Description{A conceptual framework diagram showing the progression from raw transcript data through multiple analytical layers (externalization, structuring, elaboration, and summarization) with increasing clarity and effort.}
  \label{fig:model}
\end{figure*}

\subsection{Theoretical Ground -- A Self Thought Clarification Framework}\label{sec:designrational}

We introduce a conceptual framework for self thought clarification to ground the design of \systemname. 
The framework is an instantiation 
of the \textbf{Sensemaking Model} by Pirolli and Card (Figure~\ref{fig:originalmodel} in Appendix \ref{Pirolli and Card Model})~\cite{pirolli2005sensemaking}, which was originally developed to describe how analysts interact with large datasets for sensemaking. Our formative study revealed that users face distinct challenges that align with this sensemaking process. First, static transcripts fail to support the need for flexible, hierarchical structures that should evolve as the user’s understanding deepens. Second, the process is iterative, often requiring users to add details or re-express ideas to clarify them, but users struggle to track these changes or manage AI assistance that may deviate from their intent or introduce redundancy. These findings suggest that the nature of thought clarification involves organization and iteration on one's unclear mind, which is essentially a personal sensemaking task on externalized thoughts.

The original sensemaking model contains a dual-loop process (Figure~\ref{fig:originalmodel} in Appendix \ref{Pirolli and Card Model}). The first \textit{information foraging loop} involves gathering raw data to build a foundational set of relevant information (the ``shoebox''). The second \textit{sensemaking loop} then operates on this information, involving iteratively structuring this information into a coherent schema and testing hypotheses against it.
In our context, the user is making sense of their own internal data -- their thoughts. We therefore adapt this framework 
by mapping the core stages of the sensemaking process to the comparable process of articulating and structuring one's own thoughts. 
We replace the \textit{information foraging loop} with a process of \textit{thought externalization}. 
The \textit{sensemaking loop} is slightly adapted so that the goal becomes generating perspectives through reflection instead of generating hypotheses from schematized evidence. 
While the data sensemaking process enhances \emph{structure} with effort, our process aims to improve \emph{clarity}. 


Figure~\ref{fig:model} illustrates our conceptual framework that explicates the actions and material involved in each step of self thought clarification. Our system provides four layers of functional support to facilitate each step of this process. We detail our system features in the section below.

\subsection{Key Features} \label{sec:features}
In this section, we introduce the key features at each phase of the thought clarification process that were demonstrated in the example scenario.

\begin{figure*}[t!]
  \centering
  \includegraphics[width= \textwidth]{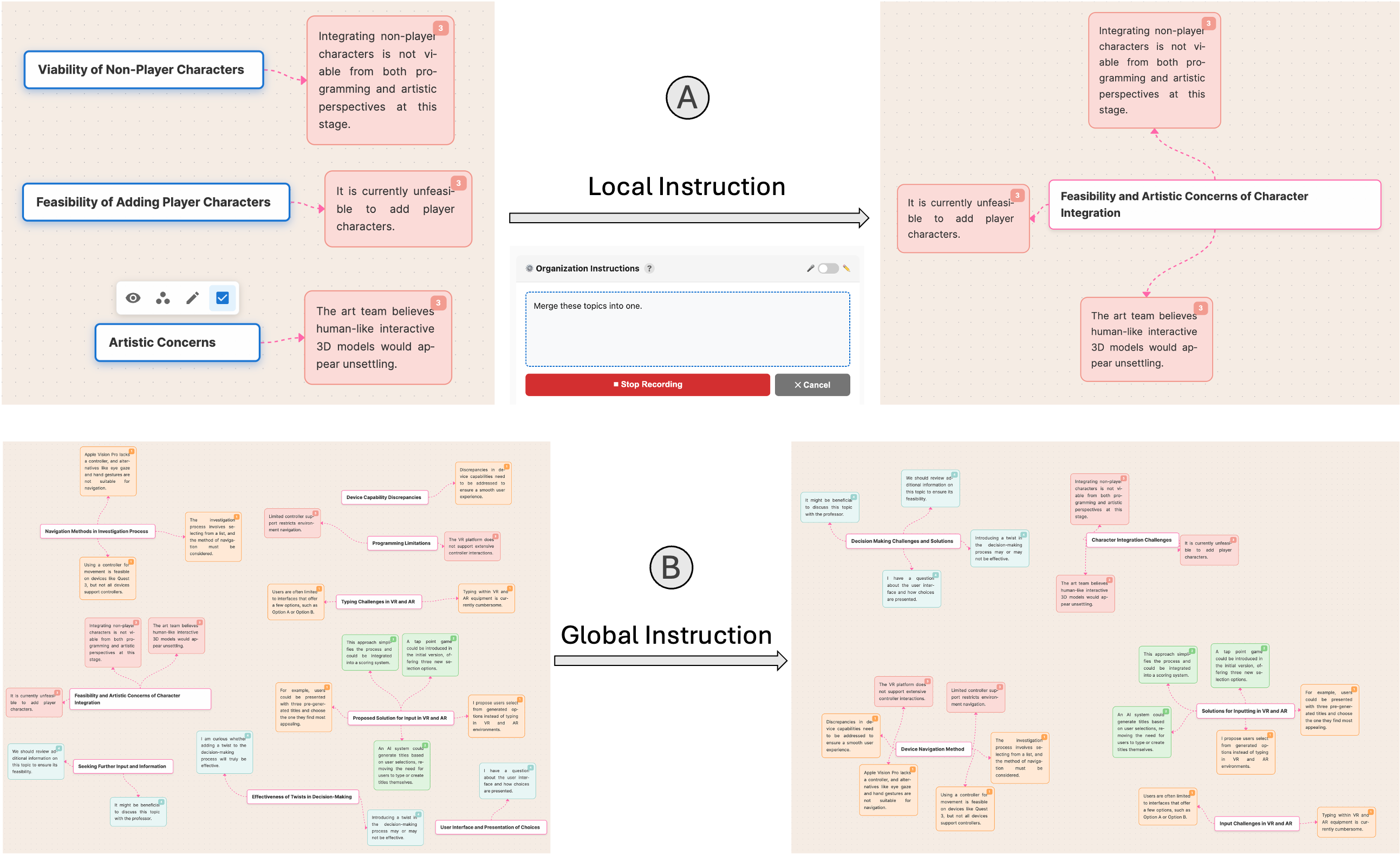}
  \caption{Verbal Restructuring: Users can either dictate local instructions that apply to selected topic groups, such as a merge request (A); or global instructions that apply to all topic groups, such as a specific new structure (B).}
  \Description{A comparison showing ``Local Instruction'' view (A) with focused node connections versus ``Global Instruction'' view (B) displaying the complete network of all interconnected ideas and concepts.}
  \label{fig:instruction}
\end{figure*}

\subsubsection{Thought Externalization Layer}
Adapted from the concept of the information foraging loop, this layer can capture the user's speech and create visual representations of the user's thoughts. This layer aims to provide assistance for the initial externalization of thought, addressing the challenges of working memory limitations and the mismatch between non-linear thought and linear speech. To achieve this, \systemname combines iterative voice dictation with a dynamic semantic space that automatically filters, extracts, and clusters ideas as they are spoken (Figure~\ref{fig:model} bottom).

\systemname uses iterative voice dictation as its primary input method. A user can speak their thoughts in a continuous stream and in multiple rounds. Directly seeing what has been spoken from a high level is essential for creating an initial mental model of the thoughts (DG1). \systemname achieves this by transforming the dictated input into a spatial and interactive Semantic Content Space (Figure~\ref{fig:teaser}). We call it a ``semantic canvas'' to reflect that the elements and their spatial organization are not arbitrary, but are algorithmically generated to represent the semantic relationships within a body of information. This 2D canvas acts as an externalized representation of the user's mental model, allowing them to identify patterns and relationships that are difficult to make sense and remember.

Our choice of a node-link diagram for the semantic space is for addressing the challenges of organizing verbal content due to its linear, unstructured, and ephemeral nature. Research on concept mapping has demonstrated that such representations lower cognitive load and support sensemaking by making the semantic relationships between ideas explicit, persistent, and malleable~\cite{novak2008theory}. Its non-linear and hierarchical representation allows users to gain an overview and inspect cross-link structures within their thoughts, while continuing to develop on threads of thinking. It also provides a flexible foundation that can represent diverse structures catering to diverse user-created topics, ranging from travel plans to persuasive arguments, making it a suitable representation for a general-purpose thought clarification tool. The technical detail of this layer is in Section~\ref{sec:pipeline} and Section~\ref{sec:algorithm}.




\begin{figure*}[t!]
  \centering
  \includegraphics[width= \textwidth]{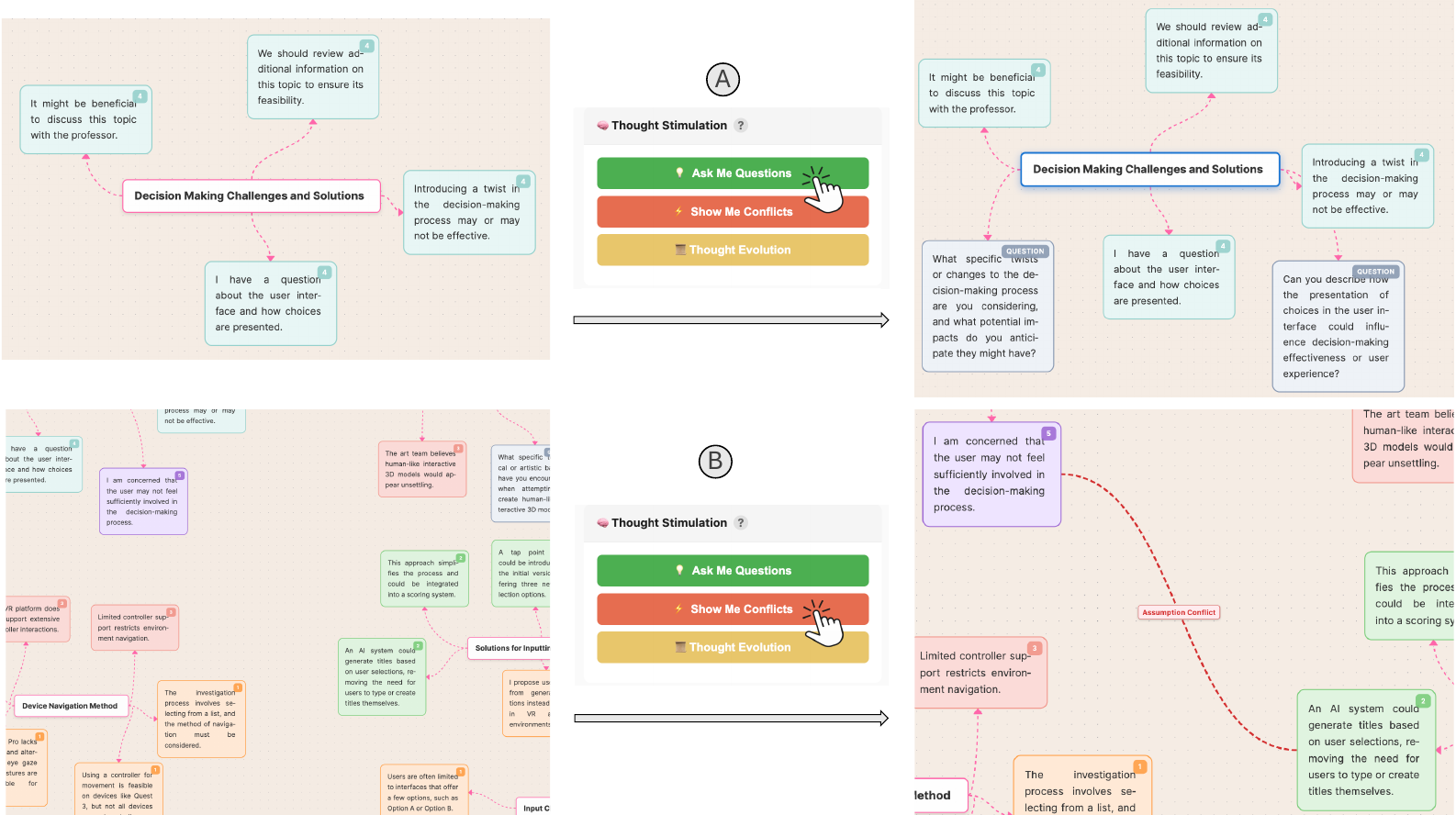}
  \caption{In-place AI Suggestions: A: \cl{For an existing topic group ``Decision Making Challenges and Solutions'' (left), user clicks the ``Ask Me Questions'' button, the system generates two guiding questions linked to the group (right). 
  B: Clicking the ``Show Me Conflicts'' button requests the system to detect logical conflicts between nodes. If found, it generates a dashed line between them and labels the conflict (right).}}
  \Description{An interface displaying an interactive thought mapping tool with connected idea nodes and three main functions: ``Ask Me Questions,'' ``Show Me Conflicts,'' and ``Thought Evolution.''}
  \label{fig:ai_suggestion}
\end{figure*}

\subsubsection{Structuring and Schematizing Layer}

The second layer aims to support users in dictating an organizational structure on their externalized thoughts (DG2). This corresponds to the top-down, hypothesis-driven activities of the sensemaking loop, where the user actively reorganizes their ideas to form a coherent schema (Figure~\ref{fig:model} lower middle). The central feature is \textit{Verbal Structuring} mechanism, which allows users to input verbal commands to restructure the semantic content space (Figure~\ref{fig:instruction}). This process enables users to efficiently test different organizational schemas without extensive manual manipulation.

The reorganization instructions can be applied at two levels of control. For broad, top-down changes, a user can input a \textit{global instruction} that applies to the entire set of nodes in the space (Figure~\ref{fig:instruction} A). For more focused adjustments, a user can first select a specific subset of topic nodes in the space and then input a \textit{local instruction} that acts only on the selected items (Figure~\ref{fig:instruction} B). The system supports several types of reorganization commands. A user can request a specific new structure by dictating the exact new topics they want, such as, ``\textit{Reorganize the selected content into three topics: `Design Goals,' `User Feedback,' and `Technical Constraints'}.'' Alternatively, a user can ask for a new structure without specifying the topics, prompting the system to analyze the content and generate a new organization based on its semantic analysis.

Furthermore, the system supports dynamic and iterative schema evolution through commands that modify the existing structure. Users can verbally instruct the system to add new topics, and the system will identify and move relevant child nodes from the transcript into the newly created topics. Users can also select multiple topic nodes and input a command to merge them into a single, consolidated topic, or select a single topic and instruct the system to split it into multiple, more granular topics. This set of verbal commands provides a fluid and low-effort method for users to refine the organization of their thoughts, directly supporting the iterative process of building and testing a mental model.

\subsubsection{Elaboration and Deepening Layer}
After thoughts have been externalized and structured, the third layer aims to help the user deepen their understanding and develop more perspectives by identifying underdeveloped areas and discovering latent conflicts (Figure~\ref{fig:model} upper middle). \systemname provides features that find the weaknesses and opportunities within their current schema, which accelerates the sensemaking loop and moves the user from organization towards a more thorough and coherent final product (DG4).

A button labeled ``Ask Me Questions'' is located in a side panel (Figure~\ref{fig:ai_suggestion} A). This feature can be applied to either the entire canvas or selected topics. When triggered with no topic selected, the system analyzes the content distribution across all topics and identifies the ones that have a comparatively low number of associated content nodes. 
It then generates guiding questions as new nodes within that topic, designed to prompt the user for further input. Alternatively, a user can first select one or more specific topics and then trigger this feature, which lets the system generate relevant questions only within the selected topics. 


A ``Show me conflicts'' button instructs the system to analyze the semantic content of all content nodes for logical inconsistencies or opposing statements (Figure~\ref{fig:ai_suggestion} B). The underlying process is designed to identify several types of logical issues, including direct contradictions (e.g., one statement claims a feature is ``essential'' while another calls it ``unnecessary''), logical inconsistencies (e.g., two statements that cannot both be true simultaneously), and strategy conflicts (e.g., proposing two mutually exclusive approaches to the same problem). When conflicts are detected, the system visualizes them by drawing edges between the conflicting nodes with the conflict types labeled on them.


\subsubsection{Reflection and Presentation} Layer
The final layer supports the final phases of the sensemaking process: re-evaluation and presentation (DG3). It allows the user to review the evolution of their thought process via an interactive timeline, fostering metacognitive awareness of how their thoughts have developed. Furthermore, it supports the final presentation stage by transforming the dynamic thought-space into a compact and coherent memo (Figure~\ref{fig:model} top). These make up the last stage of the sensemaking process, where a crystallized set of ideas is prepared for communication or future action.

\begin{figure*}[t!]
  \centering
  \includegraphics[width= \textwidth]{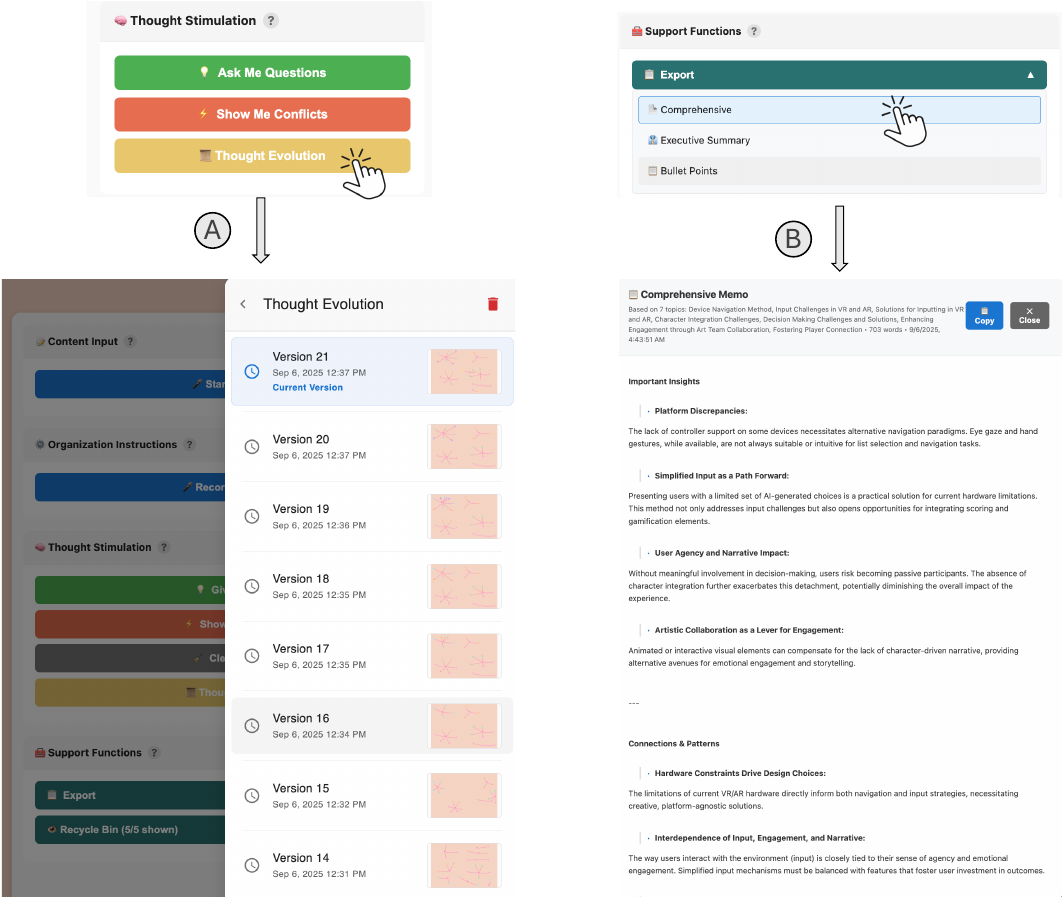}
  \caption{A: Clicking the ``Thought Evolution'' button pops up a timeline window, hovering on which will make the canvas layout go back to that stage with a smooth animated transition. B: Three types of ``Export'' formats are provided, which can generate reports with different structures for the selected canvas elements.}
  \Description{The "Thought Evolution" feature interface displays version history with timestamps on the left and export options (Comprehensive, Executive Summary, Bullet Points) on the right.}
  \label{fig:support}
\end{figure*}


A version history is automatically saved at key intervals during the user's session, such as when dictating new content or finishing manipulation of the nodes. This history is represented visually as a timeline that the user can access (Figure~\ref{fig:support} A). When the user moves their cursor to hover over a specific version point on the timeline, \systemname displays a preview of the semantic content space as it existed at that time, using a smooth animation to transition between the current state and the previewed state. If the user clicks on that version, the main interface reverts to that historical state, allowing the user to review or even resume their work from that earlier point in their process.

An ``Export'' function generates a textual summary of the content nodes selected by the user (Figure~\ref{fig:support} B). The user can choose from several formats depending on their goal. A comprehensive summary synthesizes key themes, identifies connections, and suggests potential next steps in a reflective tone. An executive summary provides a high-level overview, focusing on the most important insights and strategic implications. Finally, a bullet point summary organizes the thoughts into a structured list for quick reference. These export options provide a tangible output of the thought clarification process, suitable for documentation or sharing.

\subsection{Example Scenario} \label{sec:scenario}
In this section, we present a scenario that shows how a graduate student, Thomas, uses \systemname to organize his thoughts about an ongoing research project.


\begin{figure*}[t!]
  \centering
  \includegraphics[width= \textwidth]{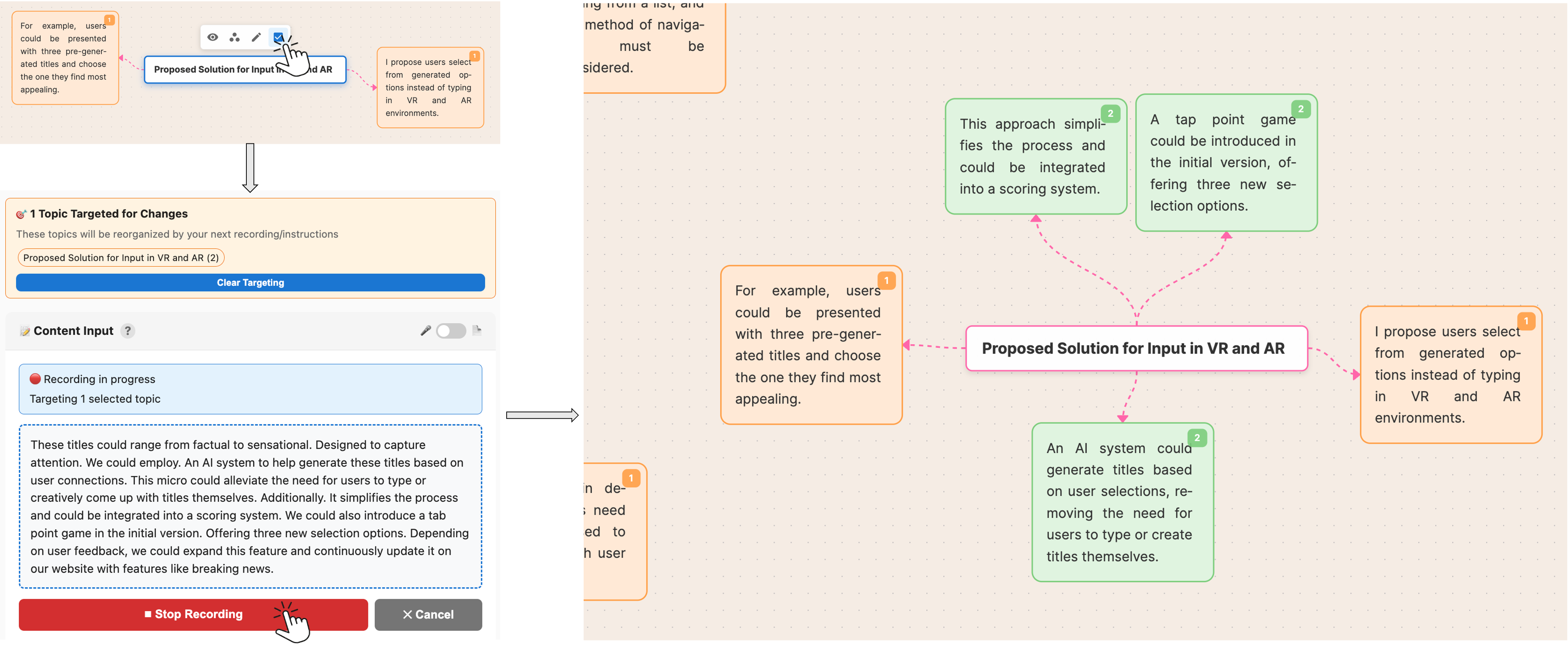}
  \caption{Iterative Verbalization: By selecting the topic group (``Proposed Solution for Input in VR and AR'') and speaking, the spoken content is analyzed and added under the selected topic group as new nodes with different colors.}
  \Description{A detailed view of the thought mapping tool during active use, showing numerous interconnected text nodes with relationship lines and a recording interface at the bottom for capturing ongoing thoughts.}
  \label{fig:selectinput}
\end{figure*}

\textit{Content Dictation and Layout Creation.} Thomas dictates his initial unstructured thoughts. \systemname extracts four key sections: ``Navigation Methods in Investigation Process'', ``Device Capability Discrepancies'', ``Typing Challenges in VR and AR'', and ``Proposed Solutions for Input in VR and AR'', displaying them as a node-link diagram (Figure~\ref{fig:teaser}). To add detail to the last topic, he selects it and speaks again; \systemname extracts the new content into that cluster, using a distinct color to indicate the second input round (Figure~\ref{fig:selectinput}).


\textit{Switching Perspectives Using a Reorganization Instruction.} To consolidate related ideas, Thomas selects multiple nodes and dictates, ``Merge these topics into one''(Figure~\ref{fig:instruction} A). Later, to view the project through a new lens, he records a global instruction with a specific new structure. \systemname reorganizes the entire information space according to these new topics (Figure~\ref{fig:instruction} B).


\textit{Thought Stimulation.} Seeking elaboration on ``Decision Making,'' Thomas uses ``Ask Me Questions'' on the target node. The system generates connected question nodes (Figure~\ref{fig:ai_suggestion} A), prompting further verbalization. He then activates ``Show Me Conflicts,'' which reveals an ``Assumption Conflict'' edge between two nodes (Figure~\ref{fig:ai_suggestion} B), prompting him to delete the flawed content.



\textit{Review and Export Thoughts.} Thomas uses ``Thought Evolution'' to scan through the timeline, previewing and reverting to previous canvas states to trace his logic (Figure~\ref{fig:support} A). Finally, he clicks ``Export'' to generate a comprehensive report summarizing key insights and next steps for documentation (Figure~\ref{fig:support} B).

\begin{figure*}[t!]
  \centering
  \includegraphics[width=0.8\textwidth]{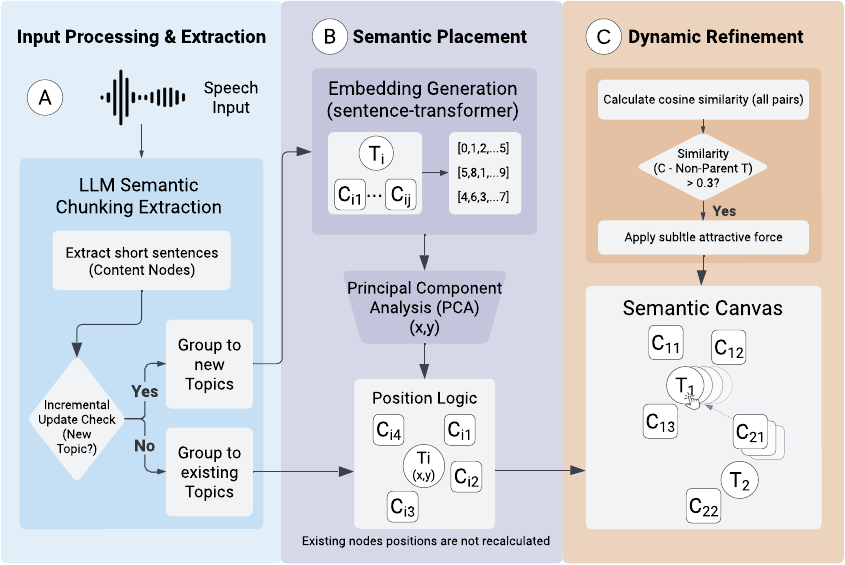}
  \caption{System architecture for speech-to-canvas processing in \systemname. The pipeline consists of three parts: (A) Input Processing extracts content nodes from speech and groups them into topics, (B) Semantic Placement uses PCA-based dimensionality reduction to position nodes spatially while preserving existing nodes' positions, and (C) Dynamic Refinement applies attractive forces between semantically similar nodes to maintain spatial coherence.}
  \Description{Three-stage pipeline diagram showing speech-to-canvas processing. Stage A (blue): speech input flows through LLM extraction to create content nodes, grouped into new or existing topics. Stage B (purple): embeddings are generated and PCA reduces dimensions to x,y coordinates while preserving existing positions. Stage C (orange): cosine similarity calculations trigger attractive forces between related nodes, resulting in a semantic canvas with spatially organized content.}
  \label{fig:pipeline}
\end{figure*}

\subsection{Implementation} \label{sec:implementation}

\systemname is implemented as a web application with a React-based frontend client and a Flask-based backend server that manages data processing, real-time communication, and interactions with LLMs. The technical details are discussed in the following sections. The prompt for each feature is shown in the Appendix~\ref{sec:prompts}.

\subsubsection{Frontend Architecture}
The user interface is developed using the \verb|React| library with \verb|TypeScript| for type safety and maintainability. The core of the interface is the semantic content space, which is rendered using the React Flow library (\verb|@xyflow/react|). This library provides a robust framework for creating node-based editors, which we use to display and manage the topic and content nodes. Client-server communication is handled through \verb|Socket.IO|, which enables real-time, bidirectional event-based communication. This is essential for features like live transcription and receiving updates from the server, such as newly generated ideas or identified conflicts. The UI components are built using the Material-UI (\verb|@mui/material|) library, which provides a consistent and responsive design system.

\subsubsection{Backend Architecture}
The backend server is built with Python using the \verb|Flask| web framework and the \verb|Flask-SocketIO| extension for real-time communication. For real-time transcription of user dictation, the server implements the AssemblyAI API, streaming audio data and receiving transcribed text in return. The core natural language processing and generation tasks are handled by calls to OpenAI's GPT-5 API, including topic extraction, question generation, conflict detection, and report generation.


\subsubsection{Semantic Chunking and Entity Extraction} \label{sec:pipeline}
As shown in Figure~\ref{fig:pipeline} A, \systemname handles unstructured, continuous speech that contains disfluencies, repetitions, and non-linear thought patterns by using LLM to perform semantic chunking extraction. To achieve this, our prompt first extracts key information as short sentences (Content Nodes), rather than keywords or short phrases, so that the nodes contain sufficient information without losing too much detail. The sentences that share the same ``topic'' (Topic Tractors) are grouped together, representing the speaker's statement or opinion. To maintain a functional balance between information density and visual clarity, we constrain the LLM to generate between one and three entities per topic per speaking turn. We determined this range empirically; allowing each topic to have more than three entities per round resulted in overly fragmented node information and was easy to cause overlapping, whereas fewer than three often led to over-abstraction that cannot guarantee the atomicity of information. When users continue speaking in subsequent rounds, the system determines whether the new input extends existing topics or introduces new ones. This allows the canvas to update incrementally, supporting the non-linear nature of spoken thought where users may elaborate on previous points across multiple turns.



\subsubsection{Two-Stage Semantic Layout} \label{sec:algorithm}
The spatial arrangement of nodes is determined by a two-stage semantic layout calculation. The first stage, \textit{Semantic Placement}, occurs in the backend after the LLM extracts topics and content from speech. As shown in Figure~\ref{fig:pipeline} B, the server first generates high-dimensional vector embeddings for each topic and content node using the \verb|sentence-transformers| library. To translate these high-dimensional relationships into a 2D layout, we apply Principal Component Analysis (PCA) using \verb|scikit-learn|. This step reduces the dimensionality of the embedding vectors to two principal components, which are then used as the initial (x, y) coordinates for each topic node. This ensures that topics with similar semantic content are initially positioned closer.

We chose PCA over non-linear dimensionality reduction algorithms, such as t-SNE or UMAP, because PCA allows almost real-time placement without the latency and is better suited for preserving global semantic distances. We calculate PCA coordinates only for the topic nodes, not the content nodes. If we included content nodes in the global PCA projection, they might drift away from their parent topics due to distinct semantic properties, which would disrupt the logical grouping of the user's statements. Therefore, content nodes are distinctively placed in a radial arrangement around their parent topic to ensure they remain visually anchored to their source.

The second stage, \textit{Dynamic Refinement}, implements a lightweight physics-based simulation to fine-tune the layout. As shown in Figure~\ref{fig:pipeline} C, this process calculates the cosine similarity between all pairs of topic and content nodes and applies a subtle attractive force between semantically related content and topic nodes if the similarity exceeds a threshold of $\tau=0.3$. We established this threshold through experimental testing. This movement provides visual cues for cross-topic connections, addressing the need for users to identify relationships between different sub-points.

\section{Evaluation}

We conducted a controlled study with 12 participants to understand the affordances of \systemname{} and evaluate its features in supporting thought clarification, by comparing it with a ChatGPT-based baseline condition. 
Through this evaluation, we aim to investigate: 
\begin{figure*}[t!]
\centering
\includegraphics[width=0.9\textwidth]{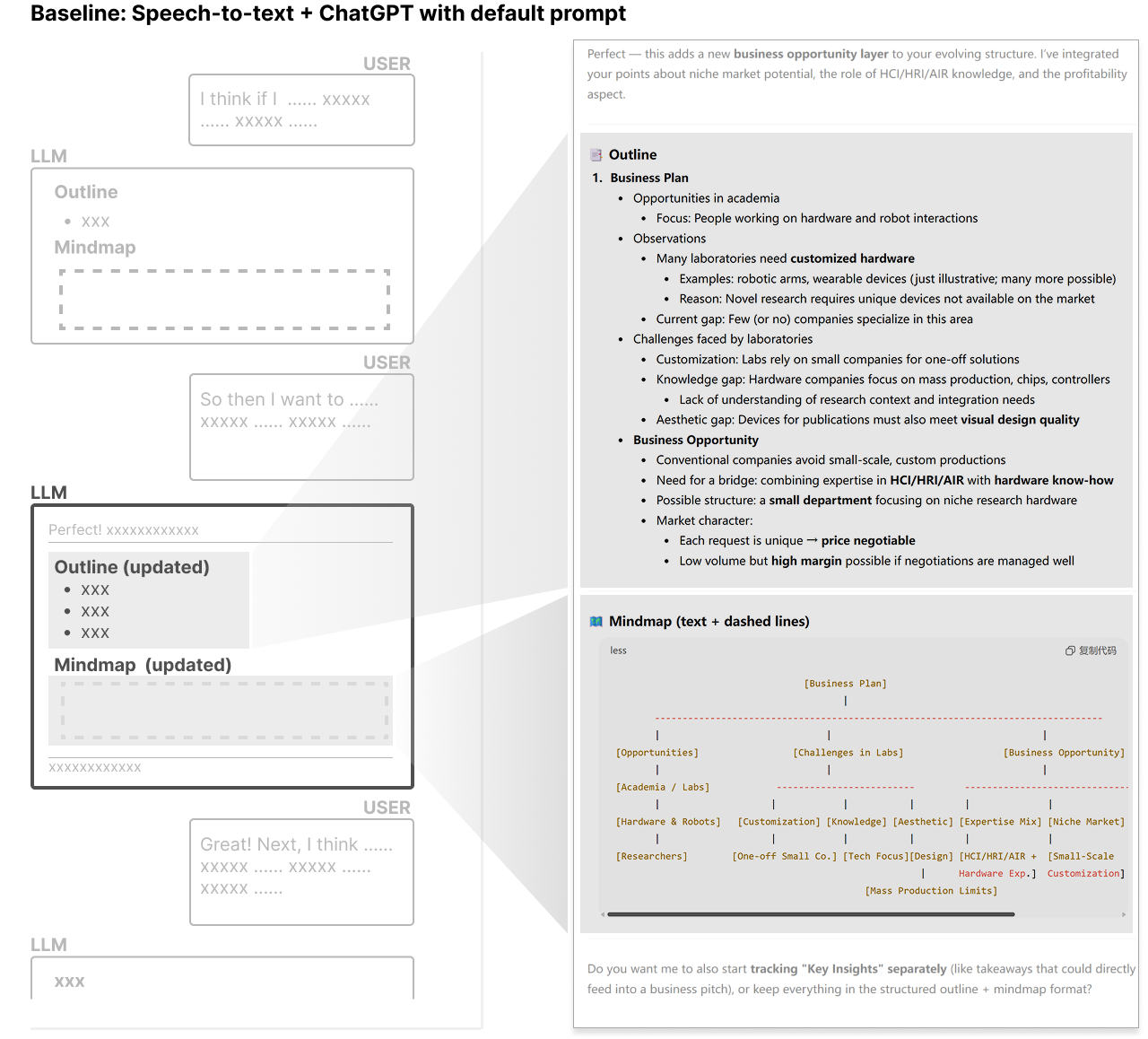}
\caption{Illustration of the Baseline interface, with a screenshot from an example conversation. To support thought organization, our default prompt asks ChatGPT to generate an outline and a text-based mindmap in each turn of the conversation.}
\label{fig:baseline interface}
\Description{This figure illustrates two different interface designs for dictation input. The left side shows the baseline interface, with a screenshot from a participant’s conversation with the LLM. The right side shows an updated interface with an outline and mind map detailing the conversation about a moneymaking plan that invites others to become business partners.}
\end{figure*}
\begin{itemize}
    \item RQ1: How did \systemname{} perform compared with the baseline?
    \item RQ2: How was each feature of \systemname{} used and perceived? 
    \item RQ3: What workflows and strategies emerged when using \systemname{}? 
    
    \item RQ4: How do \systemname{} support thought clarification compared with the baseline? 
\end{itemize}

\subsection{Baseline Condition}

The Baseline condition is the ChatGPT interface with the use of dictation input, as illustrated in Figure \ref{fig:baseline interface}, with the same setup for the formative study. It provides a state-of-the-art AI with a dialogue-based thinking partner that allows users to verbalize and organize their ideas. As participants' ability to prompt AI varies, to ensure a fair comparison between the baseline condition and \systemname{} in terms of AI support, we crafted a default prompt for ChatGPT injected at the beginning of every conversation. 
The prompt asks ChatGPT to provide a list of outline distilling key information from participants' verbal content and show a text-based concept map whenever appropriate.
Similar to \systemname{}, the baseline also helps distill and organize the verbal content, while providing a basic visual structure using a generated textual mindmap. Participants can iteratively input, organize and refine their thoughts by submitting dictated content and follow-up prompts (e.g., “help me clarify the previous content”). Participants are also allowed to use the keyboard and mouse to revisit and edit previous entries. The baseline prompt and a conversation example are attached in Appendix \ref{Baseline default promptl} and Appendix \ref{Baseline example}, respectively.

\subsection{Participants}

Twelve participants (N=12), self-reported six females and six males, were recruited via snowball sampling from a local university. 
All participants were fluent English speakers and had varied prior experience in speech-to-text technology and generative AI usage. All participants received 20 USD as compensation for their participation. Participants had varied experience with speech-to-text and generative AI. The detailed information about participants is presented in Table~\ref{demographic data}. We also report the topics proposed by each participant in Appendix \ref{sec:topic for study}.

\subsection{Task design}

Participants are instructed to conduct a ``Thought Clarification'' task using the \systemname{} and Baseline, respectively. The task was designed to elicit continuous verbalization of their thoughts, encouraging them to externalize and progressively refine their ideas with the support of the provided tools, with the emphasis placed on the gradual clarification of thinking processes rather than on generating immediate solutions. The instructions are presented in the supplementary materials.

The topics for the task were proposed by participants to ensure that both conditions could be compared fairly, by aligning participants’ familiarity and initial cognitive starting points across the two topics. Each participant was instructed to generate two distinct topics, either by developing their own or by referring to the default themes provided. The proposed topics were expected to \textit{involve a situation where they feel mentally cluttered and require clarification and development through external representation}. 
Five example topic themes are prepared as default references shown in the supplementary materials. Participants were also asked to justify their topic choices by explaining the challenges associated with clarifying them. Once approved by the experimenter, the topics were confirmed for use in subsequent tasks.

\subsection{Procedure}

This study was a 90-minute session, conducted either in-person in a usability lab or remotely, depending on participants’ preference, as all tools and procedures could be fully accessed on a laptop through the Chrome browser. The procedure consisted of three phases: \textit{Pre-study preparation}, \textit{Main study}, and \textit{Post-study interview}. In the in-person setting, sessions were held in a quiet room with a single experimenter.

\textbf{Pre-study preparation.} At the beginning, participants were required to fill out a survey about their demographic data, including age, gender, occupation, and past usage of speech input and large language models were collected. And then, they are instructed to propose two topics. When the experimenter confirmed the reported topics, they could move forward to the Main study. 

\textbf{Main study.} Participants then completed two “Thought Clarification” tasks with the two proposed topics, one in each condition. The topic assignment is randomized by the experimenter. The order of the interface condition was counterbalanced across participants. Before the task, each participant could have a practice session with the aim of getting familiar with the features of \systemname{} and Baseline, respectively. Participants were encouraged, though not obligated, to progressively refine and structure their thoughts, prioritizing depth and coherence over expediency in arriving at a conclusion. 
Participants were instructed to view the system primarily as an organizational facilitator, generate original content by themselves, and refrain from asking the system to generate a significant amount of content. 
Furthermore, in both tasks, the use of generative, open-ended directives such as “help me generate a plan” was expressly discouraged. 

Participants were recommended to finish the task in 20 minutes, but it was not a hard limit. The experimenter would remind them when time is up, and they could still continue the task until they feel their thoughts are clear enough. After each task, participants were required to rate the usefulness of the tool, satisfaction with the task, quality assessment of the features on multiple 7-point Likert scales, and complete a NASA-TLX questionnaire for rating their workload during the task. Between the two sessions, a short break is instructed to reduce fatigue effects.

\textbf{Post-study interview.} Participants were interviewed about their user experience and preference of the two conditions. To be specific, the interview protocol covered perceived advantages and disadvantages of each method, strategies adopted during drafting, sense of control over content, perceptions and use of \systemname{} features, as well as its usability, potential application scenarios, and suggestions for improvement. The detailed interview protocol is displayed in the supplementary materials. The interview lasted approximately 25 minutes.

\subsection{Data Collection and Analysis}

During the experiment, we collected multiple forms of data to capture participants’ behaviors and experiences. All task sessions were recorded via full-screen video and audio. Subjective ratings were collected using post-task questionnaires, which featured 7-point Likert scales on thought clarification satisfaction and the NASA-TLX workload scale. Additionally, we conducted semi-structured interviews after the tasks to gather participant reflections.

For the qualitative data (think-aloud protocols, transcripts, and interview responses), we employed thematic analysis to identify recurring patterns and extract key insights across conditions. Two researchers independently coded the transcripts of three participants and discussed their codes and initial themes to reach agreement. One researcher continued coding the remaining data. The final themes were iteratively defined through discussions among the researchers. 

For the quantitative data (interaction logs, video analysis, and scale ratings), we calculated descriptive statistics and performed statistical comparisons between the two conditions. 


\begin{figure}[t]
\centering
\includegraphics[width=\linewidth]{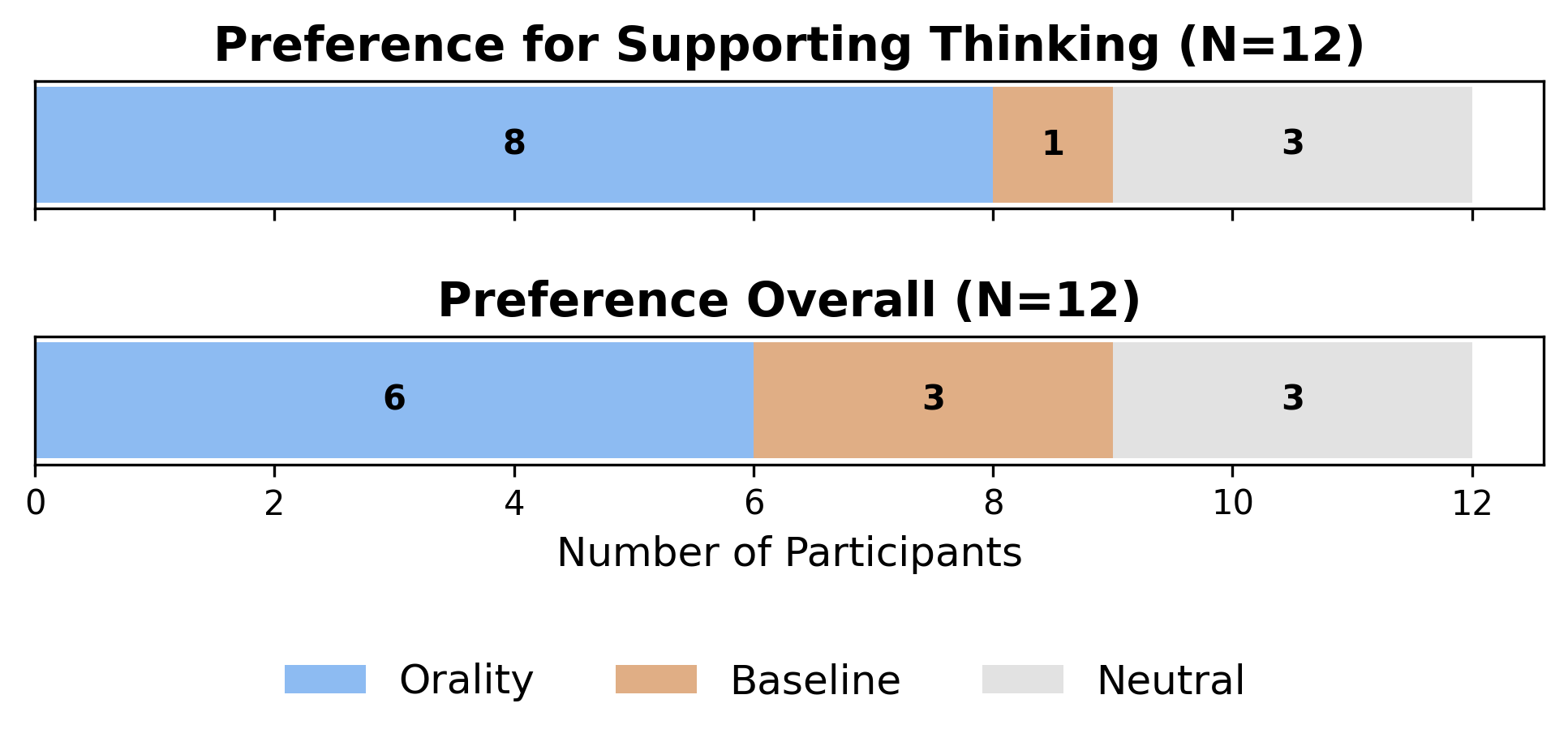}
\caption{Participants’ preferences \cl{between \systemname{} and the Baseline for supporting the thinking process and overall use, respectively.}} 
\label{fig:preference}
\Description{Two bar charts showing participant preferences strongly favoring the Orality condition over the ChatGPT+Dictation condition for thinking support and overall use.}
\end{figure}

\begin{figure*}[t]
\centering
\includegraphics[width=\textwidth]{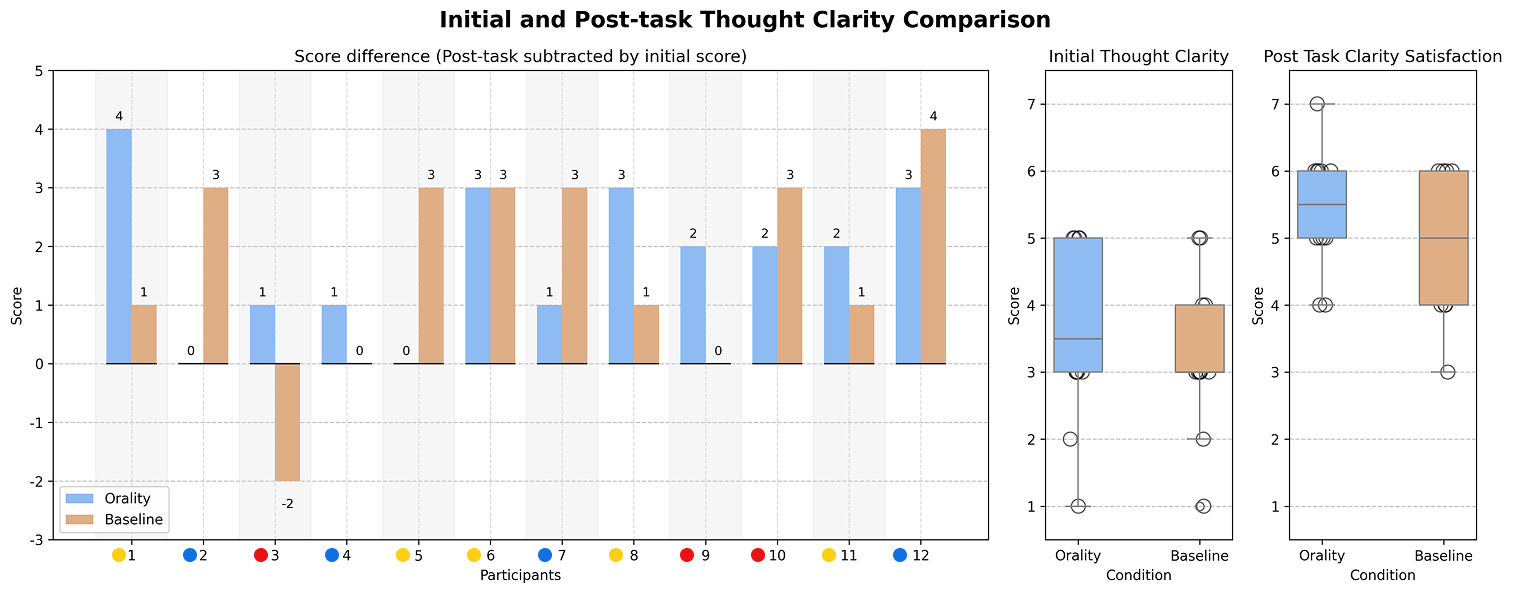}
\caption{Ratings of initial and post-task thought clarity. Left: Clarity differences between initial and post-task ratings per task per participant. 
The participants were colored by group based on their workflows identified in Section \ref{group distinguish}(yellow: Master Commanders; blue: Hands-On Mapmakers; red: Lazy Talkers). Right: Aggregated box plots of the actual clarity ratings given by all participants, before and after the task.}
\label{fig:initial thought clarity and post satisfaction}
\Description{A two-panel figure comparing initial thought clarity and post-task satisfaction ratings between two conditions, showing individual participant data and aggregated box plots.}
\end{figure*}

\section{Results}

\subsection{RQ1: How did \texorpdfstring{\systemname}{Orality} perform compared with the Baseline?} 

We attempted to measure task satisfaction in the two conditions in thought clarification with two Likert scale questions asking participants about thought clarity levels before and after the task.
As presented in Figure \ref{fig:initial thought clarity and post satisfaction}, participants rated both the initial thought clarity of Orality (M=3.58, SD=1.31) and Baseline (M=3.250, SD=1.14) and post-task clarity satisfaction of Orality (M=5.42, SD=0.90) and Baseline (M=4.917, SD=0.996). No statistically significant difference was found between the two interface conditions by comparing the differences in clarity before and after the task, although the \systemname{} condition showed a tendency toward higher clarity. We observed some interesting differences among participants who adopt different workflows, which we will explain in Section 6.3.

Participants were asked about their preference for using the tools. 
As shown in Figure \ref{fig:preference}, eight out of twelve participants chose \systemname{} \cl{when being asked which tool could better support their thinking process, while one chose the baseline. Interestingly, when being asked for their overall preference as which one they liked better, six preferred \systemname{} while three preferred the Baseline. This difference could be partially explained by the different thinking goals participants had for their tasks, see more in Section~\ref{active_sensemaking}.} 
Three participants reported no preference for both measures. Some participants noted that the Baseline was more suitable for certain specific topics. For instance, P3 preferred the Baseline when addressing a specific question, but favored \systemname{} when thinking from a global perspective on a general problem. P7 appreciated that \systemname{} provided a structure for transforming general thoughts into specific ones, while the Baseline was more suitable for small-scale, small-talk tasks. P12 considered the Baseline more suitable for achieving a specific goal, whereas \systemname{} was better suited for broader discussions such as brainstorming.


We found no statistically significant differences between \systemname{} and the Baseline for subjective workload (NASA-TLX). The lack of a statistically significant difference in our quantitative measures is not surprising, given that the experimental task is an open-ended, participant-proposed thinking task for maintaining external validity. The choice of task and participants' thinking strategies differ widely across tasks. We continue our comparison between \systemname{} and the Baseline in Section~\ref{qualitative} with qualitative analysis, which yields a more insightful comparison.

\subsection{RQ2: How was each feature of \texorpdfstring{\systemname}{Orality} used and perceived?}

\subsubsection{Quality Assessment and Usefulness}
\begin{figure}[t]
  \centering
  \includegraphics[width= 0.5\textwidth]{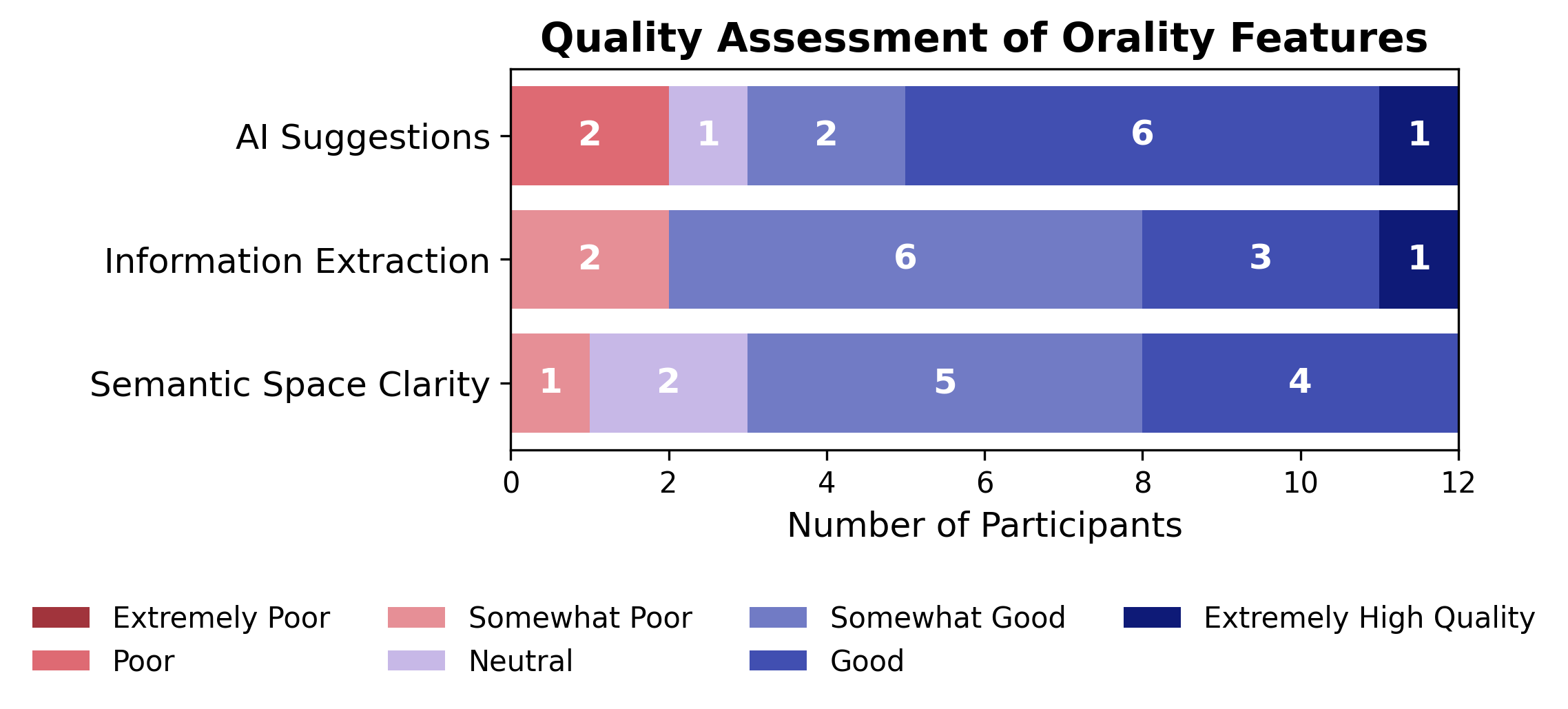}
  \caption{Participants' assessment of \systemname{} features' output quality. ``Information Extraction'' and ``Semantic Space Clarity'' refer to system's ability to identify topics from user speech and represent them visually. ``AI Suggestions'' refer to the relevance and quality of AI-generated questions and detected conflicts.}
  \Description{A stacked horizontal bar chart showing participant quality ratings for three Orality features (AI Suggestions, Information Extraction, and Semantic Space Clarity) on a seven-point scale from "Extremely Poor" to "Extremely High Quality," with most ratings falling in the positive range (somewhat good to extremely high quality) across all three dimensions.}
  \label{fig:quality-assessment}
\end{figure}

\begin{figure}[t]
  \centering
  \includegraphics[width= 0.5\textwidth]{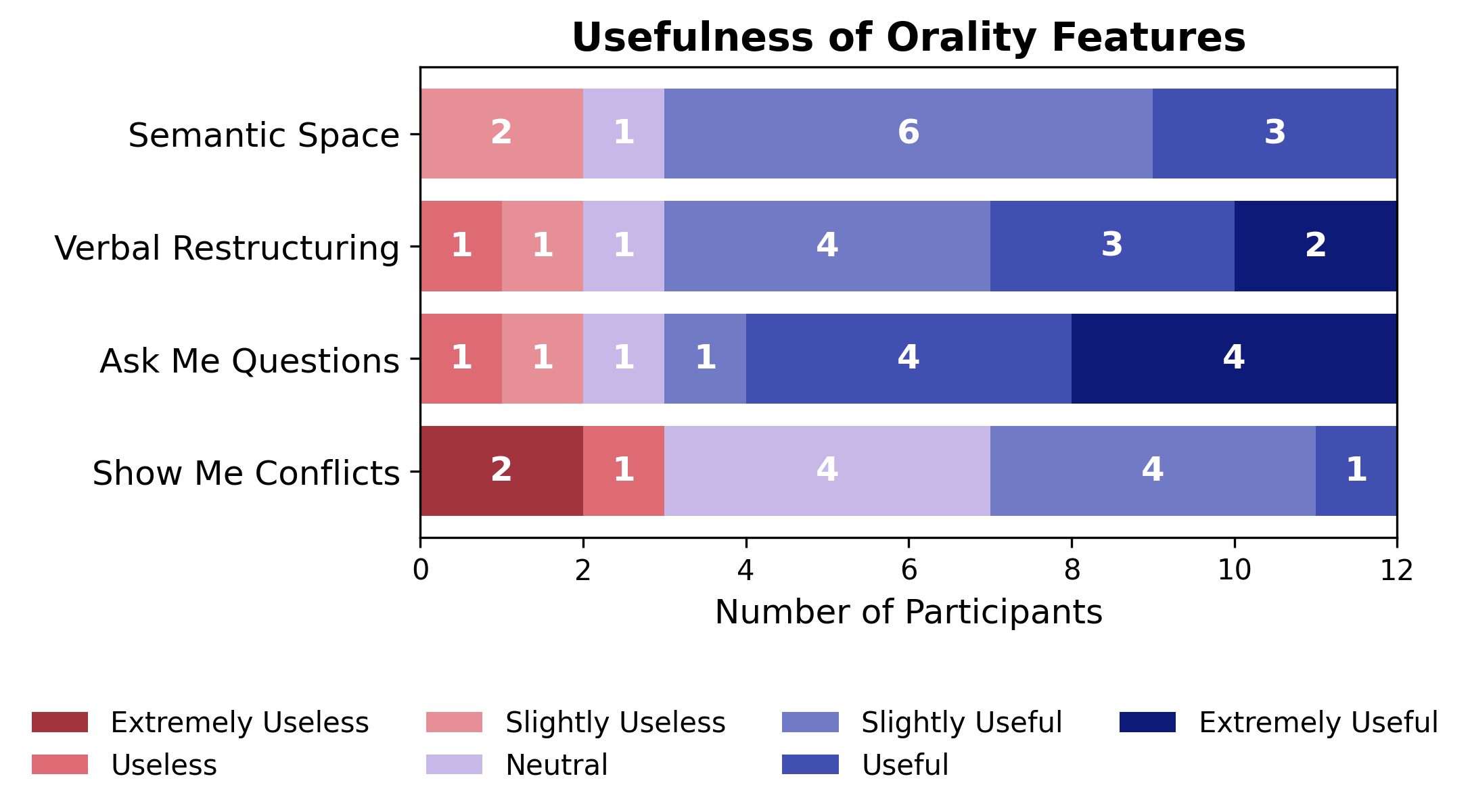}
  \caption{Participants' assessment of usefulness for \systemname{} features.} 
  \Description{A stacked horizontal bar chart showing user ratings of four Orality features ("Semantic Space," "Verbal Structuring," "Show Me Ideas," and "Show Me Conflicts") on a usefulness scale, with most participants rating the features as useful to extremely useful.}
  \label{fig:usefulness}
\end{figure}

We calculated the descriptive statistics for each quality scale (Figure \ref{fig:quality-assessment}) as a reference: information extraction of spoken contents (M = 5.08, SD = 1.16), Semantic Canvas Clarity (M = 5.00, SD = 0.95), and AI suggestions (M = 5.08, SD = 1.62). We can see the main output of the system has a reasonable performance overall while individual cases vary.
The surveys also assessed the usefulness of each \systemname{} feature (Figure \ref{fig:usefulness}). While the majority of ratings are positively rated from ``slightly useful'' to ``extremely useful'', individual differences are also large. Ask Me Questions received the most positive ratings, followed by Verbal Structuring and Semantic Space. Show Me Conflicts seemed to be less appreciated. In the interview, while most participants acknowledged their value, several issues were reported. 

\subsubsection{AI suggestions for thought stimulation} \label{sec:suggestions}

\paragraph{\textit{Ask Me Questions} stimulates new and deeper thoughts with in-place cues.} 
Almost all participants valued the AI-generated inspirational questions for supporting their thinking in different situations. Many reported that suggestions were particularly useful when they did not know what to say next. For example, P7 was surprised that a casual remark triggered an unexpected question from the system that inspired a new idea, while P3 noted it helped unblock their thoughts when unsure what to say. Additionally, participants also found it helpful for elaborating on abstract topics: P2 explained that the system prompted him to provide concrete examples, P4 compared it to breaking ideas down ``\textit{like branches on a tree},'' and P6 used it to expand subtopics. P7 further expressed appreciation for 
the question-format suggestions, rather than statements, opened new directions and encouraged deeper reflection, such as ``\textit{What is my purpose}?''
In the initial phase of the tasks, participants tended to use it to receive inspirational feedback immediately after articulating their first ideas; for example, P9 explained that early questions helped clarify initially vague thoughts. In the middle phase, participants like P12 preferred to use the feature once they had already developed some familiarity with their ideas, finding the prompts more effective at that point.

\paragraph{\textit{Show Me Conflicts} can be useful but needs improvement.} 
P12 highlighted its value in identifying areas of overlap or contradiction between two possible paths, while P3 suggested that conflict detection should be automated by the system rather than manually initiated. P8 struggled with the lack of support for resolving conflict after it's detected. P4 had doubts about the system when it did not detect a conflict they identified by themselves -- revealing ambiguous interpretations about what a ``\textit{conflict}'' is.

\subsubsection{\textit{\verbalFramework}} \label{sec:verbalframework}
This feature was reported as useful for organizing their thoughts after seeing the initial clusters generated by the system. It provides users with the freedom to verbally reconstruct their ideas to see different perspectives of their thought processes, while issuing advanced manipulation operations. An analysis of their commands reveals four distinct use patterns.


The most common pattern was for \textit{specific, high-level organization}. Participants used this to impose a clear, thematic structure on unstructured content. For example, P5 instructed the system to ``\textit{Organize my notes into three parts: requirements, framework information, and relevant information,}'' while P7 commanded a similar top-down reorganization into ``\textit{background, challenges, and approaches}.'' 


Participants also used \textit{merging or splitting} commands to control the granularity of the information display. 
Rather than defining a rigid new structure, these instructions tasked the AI with improving the existing one. This involved consolidating information, such as when P10 asked the tool to ``\textit{Combine some of my ideas to make the graph clearer,}'' or when P3 requested it to ``\textit{For the `relevant information' section, divide it by different task types.}'' As P1 explained, when facing excessive categories, using a command to combine them gave him a new perspective and a deeper understanding of his own ideas. P10 commanded it to split a parent node into child nodes when the initial layout did not match their mental model.

A third pattern involved using the system for \textit{synthesis to answer specific questions}. In these instances, users treated the system as an analytical partner capable of interpreting their own data. For example, after dictating their professional history, P2 asked the system to ``\textit{give me ideas on why I'm a good fit for the user researcher role,}'' which required the tool to find and group relevant nodes to provide a visual answer.

Finally, participants used \textit{\verbalFramework} for \textit{advanced analytical and formatting tasks}, pushing the boundaries of the feature's designed capabilities. Instructions like P12's request to ``\textit{Find the... conflict points between the paths}'' or P8's command to ``\textit{Summarize four projects under a main topic}'' show a desire for the system to perform higher-order analysis. 

\subsubsection{\textit{Semantic Canvas for Thought Clarification and Development}}

\paragraph{Helping Monitor and Develop Thoughts} 
Participants emphasized that extracting topics from spoken input helped reduce redundancy and repetition, while grouping related nodes by color and position created a structured canvas that served as thought blocks. P1 described it as ``\textit{keeping the mind clearly on the canvas},'' which helped him recall earlier content during later stages. Similarly, P11 noted that arranging nodes neatly allowed him to gauge the clarity of ideas: 
more branches indicated greater clarity, whereas fewer prompted him to refine ideas further using follow-up questions.

\paragraph{Personalized Meanings of Spatial Information}
Participants actively manipulated the \textit{Semantic Space} to inspect relationships between thought blocks. They used spatial information to represent relationships and priorities. For instance, P12 divided the canvas into left and right areas to compare clusters, while P9 placed related nodes close together or arranged subtopics around a central node. P9 also prioritized nodes by placing important ones at the center, supplementary content below, and less important ones toward the margins.

\subsubsection{\textit{Thought Evolution} for Exploring Alternatives}
This feature was used by participants to revisit and restore earlier versions. Seven of twelve participants used it to revisit previous versions, which helped them retrieve forgotten threads, compare structures across versions, or recover deleted content. P5 explained that reviewing earlier layouts allowed him to recall initial versions and identify changes, while P11 and P12 emphasized its value as an \textit{Undo} function when newly generated layouts were unsatisfactory or contained errors. Participants also reported that the ability to review and revert changes encouraged them to try new structures without the fear of losing valuable content or disrupting their logic. For example, P2 utilized this function multiple times within a short window while refining their interview responses. By restoring previous states, the participant was able to test different narrative structures for their work experience before settling on a final version. P8 described using this workflow to iterate on new directions: they would speak a new topic, assess if it was ``\textit{particularly good},'' and if not, return to the previous version to think about how to improve it. Beyond trial-and-error, the contextual review supported metacognitive monitoring of logical flows. P9 explained that reviewing the history helped ``\textit{clarify the entire logical flow},'' allowing them to see how their logic had organized up to that point.

\subsubsection{\textit{Export} for a Complete Answer}

Finally, only a few participants used and appreciated the \textit{Export} feature as a way to summarize their process of thought clarification into a textual report, which they viewed as a useful outcome of their interaction. P1 noted that it was ``\textit{more like documenting my previous thoughts},'' while P10 described it as ``\textit{something I can use later as a memo}.'' Others emphasized its value for clarifying fragmented ideas after multiple iterations, with P8 stating that the report ``\textit{helped me once to go through everything at once and make the ideas clearer}.''

\subsection{RQ3: What workflows and strategies emerged when using \texorpdfstring{\systemname}{Orality}?}
\label{group distinguish}

Analysis of the observational data (system logs, screen recordings) revealed three distinct patterns of the interaction with \systemname. This section describes them with supplementary quotes from participants' own descriptions and explanations. We categorize the participants' workflows by their use of input modalities, interaction patterns, and their underlying mental models for human-AI thought organization. As the participants' task strategies are inherently affected by their personal style and choice,  we synthesize and report them by grouping our participants into three groups 
\textit{Master Commanders}, \textit{Hands-on Mapmakers}, and \textit{Lazy Talkers}. A list of detailed descriptions of the workflow by each participant is attached in the Appendix~\ref{sec:workflows}.

\begin{figure}[t!]
  \centering
  \includegraphics[width= \linewidth]{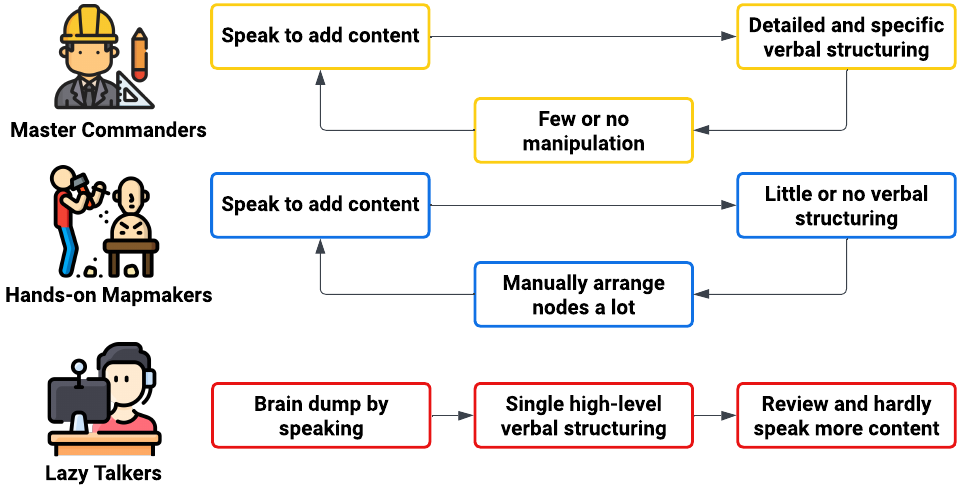}
  \caption{Three observed interaction strategies and task workflows using \systemname{}.}
  \Description{A workflow comparison diagram showing three user types—master commanders, hands-on mapmakers, and lazy talkers—each with distinct approaches to verbal content creation, ranging from highly structured to free-form brain dumping methods.}
  \label{fig:workflows}
\end{figure}

\subsubsection{Master Commanders (P2, P5, P7, P8, P12)}

The first group and the largest group operated almost exclusively in both content dictation and a series of specific, often complex, voice instructions (Figure~\ref{fig:workflows} Top). The signature behavior for this group was using a large number of \textit{\verbalFramework} instructions throughout the session, with none or minimal manual manipulation. They used a sequence of instructions to achieve their goals and also heavily used thought-guided questions. They were not just organizing content; they were actively conversing 
with the AI canvas to structure, synthesize, and even analyze their ideas.

For example, P5 initiated a session by dictating a list of research frameworks and task types and when the initial reorganization was not fully accurate, instead of manually sorting these nodes, \wx{P5 used a structural command to define the map: ``\textit{I want to organize my notes in three parts. The first one is the requirements... And the third one is all the framework}.'' Later in the session, P5 used two other instructions: ``\textit{Relevance information is too dense, divide it by different task types in each node}'' and ``\textit{combine the repeated information}.''} This demonstrates a workflow where the user treats the AI canvas as an organizational assistant that can be guided through iterative verbal commands. \wx{Similarly, P2 used a series of \verbalFramework to organize their academic and professional history into clear categories (e.g., ``\textit{Research Assistant Experience},'' ``\textit{Professional Work Experience}''). The process of P7 was notably hierarchical: first, they commanded a broad, top-down reorganization of all content into an analytical framework (e.g., ``\textit{Background},'' ``\textit{Challenges},'' ``\textit{Approaches}''). Then, P7 ``\textit{zoomed in},'' selecting a single new section and issuing another granular command to restructure its internal layout.}

These users chose this workflow because they believe that voice is a more powerful, accurate, and efficient medium than direct manipulation for structuring complex thoughts. They organized content almost exclusively through \textit{\verbalFramework} because it allowed them to perform structural and analytical tasks that would be complicated or impossible by dragging and editing the nodes. This is evident in the log of P12, who not only organized topics spatially but also asked the system to perform a logical analysis. They instructed the system to ``\textit{find the interruption part... overlap part... and conflict part}'' between their two career paths. 
Later in the interview, P12 explained that
this is because their analytical goal required it, noting, ``\textit{I asked him to list the overlapping parts between the two paths... I wanted to clearly see which parts overlapped.}''

\subsubsection{Hands-on Mapmakers (P1, P6, P9, P11)}

The second group demonstrated a workflow that relied on a tight integration of speech and direct manipulation (Figure~\ref{fig:workflows} Middle). Their signature behavior included speech in short, iterative bursts, followed by a dominant phase of manual reorganization to create spatial meaning. They used \textit{\verbalFramework} at a very low rate, preferring to structure their thoughts manually. 

For example, P9 treated the tool primarily as a visual project planning canvas. \wx{After a comprehensive initial introduction of project components, P9 engaged in meticulous manual arrangement, executing a sequence of seven consecutive topic drag events to organize clusters such as ``\textit{Project Compliance},'' ``\textit{Collaboration},'' and ``\textit{Immediate Project Requirements}'' to three corners of the space.} P9 described this arrangement process ``\textit{When I first arrange things, I start by prioritizing the most important keywords, then arranging them from top to bottom...I move the small bubbles of the major branches to the sides for easier reading.}'' \wx{Similarly, after each small addition of content, P1 spent much time manually arranging the nodes, using the spatial layout to build connections and structure their thoughts.} P1 explained their step-by-step method for manual categorization: ``\textit{I first separated them, then read each one to see which was newly generated and which was later. After reading this, I decided to further categorize the bubbles based on my own ideas}.''

The choice to adopt this workflow was rooted in the belief that thinking is a tangible, visual, and spatial activity. These users did not fully trust the AI's logic to represent their thoughts, or they found it more interesting to manually operate, so they relied on direct manipulation to physically shape their ideas and create meaning through arrangement. As P9 explained: ``\textit{I think arrangement aids understanding. Because there might be some initial issues... I might first place them alongside the initial input, comparing them... I find the process of arranging things quite interesting.}'' They found automatic layouts confusing or restrictive: ``\textit{I asked him to help me reorganize and delete the duplicate content, and then I found that only three points were left, ... made me feel a little insecure}.''(P9) and ``\textit{I wanted it to do this because there was still a lot of overlap,... it didn't seem to work}.'' (P11) For these users, the process of sculpting the canvas was inseparable from the process of understanding.

\subsubsection{Lazy Talkers (P3, P4, P10)}
The last group used a workflow characterized by a linear, two-step process: speak everything, then command a little (Figure~\ref{fig:workflows} Bottom). Their behavioral signature was a high volume of speech delivered at the beginning of the session, like a ``\textit{brain dump},'' with minimal to no manual manipulation. 

This pattern is clearly identified by P10, who initiated their session with three consecutive, long inputs covering their career dilemma, the pros and cons of a PhD versus work, etc., before 
issuing one single, global instruction: ``\textit{Combine some of my ideas to make the graph clearer}.'' 
P4 similarly demonstrates this behavior. 
After a long ``\textit{brain dump}'' regarding all facets of their current life and career uncertainty, they asked the AI to structure the conversation to focus specifically on ``\textit{post graduation uncertainty}.'' \wx{Also, P3 began by verbally outlining a clear problem (a rejected academic paper) and its context. They then further elaborated by purely speaking,
to break down the specific reviewer feedback and brainstorm potential solutions.} The long initial ``\textit{dump}'' phase was also described by P2, who felt verbalization itself was effective: ``\textit{I just wanted to describe myself first... I talked about everything that came to my mind first, and then I felt that I had partially reached my goal}.''

The interviews showed that this behavior stems from a clear cost-benefit analysis. These users found the process of manual organization to be ``\textit{expensive},'' and they felt the visual layout was irrelevant to their final goals. P10 was explicit about the high effort of manual layout driving their choice: ``\textit{I think the layout method is quite expensive to use... I felt it was too cumbersome to arrange, so I simply did not do it. But overall, I can understand the hierarchy it is trying to convey.}'' P4 explained that they disregarded manipulation because the visual layout was irrelevant to their goal, stating, ``\textit{I read them all, but I did not arrange them because I felt they were all connected by a node. I did not think it would make any difference whether they were arranged or not}.''

\subsection{RQ4: How do \texorpdfstring{\systemname}{Orality} support thought clarification compared with the Baseline?}  \label{qualitative}
This section reports our findings on how our tested tools supported participants' thinking processes for reaching their individual goals, as emerged from our thematic analysis of the interview data.


\subsubsection{Supporting divergent thinking: guiding continuous expansion vs. supplying extra information}
Our analysis showed that both \systemname{} and the Baseline support divergent thinking, but in different ways. \systemname{} was found to scaffold users' thinking by guiding them to think extra, while the Baseline was appreciated for supplying external knowledge and letting thoughts flow freely. P5 described how \systemname{} provided a ``\textit{scaffold}'' by extracting topics and providing AI suggestions: ``\textit{... it guides me to think about the categorization, even though it may be wrong, I can iterate by providing instructions. Then it also gives me the AI suggestions and detect conflict, they guide me to think about extra ideas, I find this helpful}'' P9 also said the questions asked by \systemname{} ``\textit{helped me jump out of what I have in my brain to expand my thinking.}'' 
On the other hand, the supply of external information by the Baseline was appreciated. P2 liked the Baseline's supplement of additional ways for international students to save money while brainstorming about this topic. 
P5 was making a travel plan with the Baseline, which added information on accommodation and food for them. P5 commented, ``\textit{(Baseline) gave me some extra information which helped me input more to complete the plan.}'' P1 articulated their different mental models when using the two systems: ``\textit{(\systemname{}) makes you solve every small problem thus gradually solve the big problem, it's a problem-solving tool. Working with the Baseline is more like brainstorming, you keep thinking non-stop and the system summarizes for you.}'' Such a free-thinking state was also commented by P6: ``\textit{With (\systemname{}) I can follow its topics to speak, whereas the (Baseline) didn't organize it much, might be more free.}''


\subsubsection{Supporting sensemaking: encouraging active sensemaking vs. providing answers} \label{active_sensemaking}
First, both systems support sensemaking through the act of speaking. As P9 expressed ``\textit{When speaking, my brain already went through it once... The act of speaking helped me organize what I needed to do because I had to describe it.}'' Afterwards, the most salient difference observed is that \systemname{} guides participants to come to their own conclusion while the Baseline helps by providing answers. This is well articulated by P7: ``\textit{(Baseline) leans to extract the essence, while (\systemname{}) grows a complete framework. By `essence' I meant the two takeaways, exactly what I needed at the end.}'' P7 also said the framework created with \systemname{} guided them ``\textit{from vague to concrete}''. Based on these, P7 concluded that the Baseline helps better with a smaller or more subjective idea because it directly nails down the core, while \systemname{} suits for heavier content.
Similarly, P6 said, ``\textit{(Orality) is suitable to be used for analyzing, while (Baseline) organizes thoughts and gives me its opinion.}'' Perhaps this led to a behavior pattern noticed by P9 himself, that they were mostly ``\textit{correcting}'' the Baseline by telling ``\textit{what was wrong}'', whereas with \systemname{} they were ``\textit{expanding descriptions}'' based on the AI-generated questions.

Furthermore, some participants reported that they iterated more with \systemname{} than the Baseline. P4 contrasted the differences using these two tools: \systemname{} ``\textit{stimulated me to speak more piece by piece}'' so the results are more ``\textit{complete}''; Baseline ``\textit{completely depends on how much I can squeeze out from my head}.'' P3 supported this, observing that they tended to ``\textit{say more at once}'' with the Baseline, whereas the system supported an ``\textit{iterative process}.'' 

Although the Baseline also offers mindmaps to summarize and structure users' content, the lack of interactivity and visual continuity appears to be its main drawback. The chat-based mindmaps, though basic, provided hierarchical structures that some participants found helpful. P6 described it as a ``\textit{semi-finished product}'' that nonetheless supported organization, and P10 appreciated its multi-layered expansion of content. P2 was frustrated by not being able to fix some conflicts in the Baseline's concept map by prompting, despite finding it helpful for categorization and information completion. P9 said they were more tolerant of inaccuracies of \systemname{} than the concept map in Baseline because it cannot be edited easily. P1 expressed  preference for the semantic canvas in \systemname{} because it ``\textit{guides me to think deeper}'' but the mindmap in the Baseline ``\textit{felt more like a summary}'' and didn't ``\textit{gave me directions to think}.''

However, participants had mixed feelings about more active sensemaking depending on their task goals and preferences \cl{(see Appendix~\ref{sec:topic for study} for lists of all the task topics).} Participants preferred the Baseline when they wanted AI to directly provide solutions or improve their outcome. For instance, P3 was seeking solutions to a problem, did not find \systemname{} useful because ``\textit{If I were doing content analysis then categorizing into themes would be helpful. But today I was describing my difficulties, so the topics weren't very meaningful.}'' P12 appreciated that the Baseline directly ``\textit{refined my idea well}.'' P2 also preferred using the Baseline because it ``\textit{lowered mental workload}'' and said that \systemname{} made them ``\textit{think more}'' because ``\textit{I needed to judge whether the categorization made sense and think about how to optimize it}.'' P4 also did not like the fact that \systemname{} did not help them ``\textit{converge},'' felt it was more like an ``\textit{analyzer}'' yet what they wanted were ideas.

\subsubsection{Supporting in-depth thinking: continuous diving-in vs. same-level jumping}
\systemname{} was reported to help break down big problems and facilitate ``vertical'' deepening of thoughts. Participants contrasted this with the Baseline, noting that standard LLM interactions often produced ``horizontal'' summaries that listed parallel concepts without depth.  P1 described that while the Baseline kept their thinking at a single level, such as comparing the general pros and cons of cities, \systemname could guide them to break a high-level goal like ``\textit{how to be a software architect}'' into specific actionable steps, such as ``\textit{backend development}'' to ``\textit{server development}'' to ``\textit{specific languages},'' and finally into specific learning resources.

Participants reported that \systemname{} supported a smoother thinking process, as the visual layout enabled more focused thinking by helping users track what had been externalized and plan next steps. When their thinking was interrupted, participants could resume by elaborating on clusters that lacked content (P1, P10) or by continuing with subtopics of interest (P6). 
By contrast, the Baseline sometimes led to interruptions for thinking, with participants experiencing confusion when conversations became long and text-heavy (P1). This required users to expend additional effort parsing and extracting key points from dense history (P4, P7, P8), as well as identifying new updates compared to previous exchanges (P4).


\subsubsection{Supporting thought development: targeted thinking vs. partner-like chatting}
\systemname{} was reported to have effective support for focused and targeted thought development, particularly because of its topic extraction and node-selection feature. 
P1 described that the initial clusters of their spoken content ``\textit{kept my focus on those two themes}'' and felt their subsequent thought expansion was based on them. 
P6 echoed this by saying they ``\textit{spoke surrounding a topic}'' with \systemname{}, whereas with the Baseline it was ``\textit{more divergent}.''

Being able to apply operations on a selection of nodes helped participants to develop targeted areas with more control. P8 emphasized the importance of this so that they could ``\textit{think or ask questions referring to a target, ... It helps to think about one thing in several parts, or use it as context to ask for system suggestions},'' while contrasting this with the Baseline as being ``\textit{messy on both my end and the system's end}.'' P6 capitalized this feature by enforcing an order of a sequence of AI operations. They used the ``\textit{Ask Me Question}'' function on every topic first before checking potential conflict with ``\textit{Show Me Conflict}''. P10 felt \systemname{} can better support thinking exactly because of this: ``\textit{It asks questions based on specific nodes, this makes the layers clear. But if (Baseline) asks me questions, I wouldn't know that it refers to}.''

On the other hand, certain engagement effects rest in the Baseline's chat-based interface. Some participants valued the ``\textit{partner-like}'' feeling that provides validation. As P1 described that when their ideas of next steps aligned with the questions asked by the Baseline, they felt more confident about it, ``\textit{because this back-and-forth discussion makes you feel like discussing with a partner, it responds to me, making me feel more certain about the next steps}.''

\subsection{User-suggested improvements and potential use scenarios}

\paragraph{Addressing information overload with excessive nodes.} Participants reported that an excessive number of nodes led to visual clutter and cognitive overload. They proposed solutions focused on both interface design and automation. Interface suggestions included predefined layout templates, such as tree structures, to organize nodes systematically (P4). To reduce informational density, P9 recommended automatically extracting and emphasizing keywords to support faster scanning. On the other hand, automation was also proposed for real-time merging and categorizing. For example, P1 wanted automatic sorting into user-defined groups, while P3 suggested clustering similar content during input to prevent spatial overcrowding.


\paragraph{Providing more node interactions.} Participants also requested more flexible and semantically meaningful node interactions. They wanted to create or modify relationships between different nodes, such as visually linking design examples to theoretical frameworks (P5). Another key suggestion was semantic zoom, which would allow users to collapse specific nodes (e.g., ``\textit{advantage of} city X'' and ``\textit{advantage of} city Y'') into a broader parent category like ``\textit{advantage of a city}'' (P6). Finally, users wanted to group nodes from multiple perspectives beyond the current time-based color coding. For instance, P7 suggested regrouping content from ``\textit{approaches}'' or ``\textit{challenges}'' into new categories like advantages or limitations.


\paragraph{Suggested potential user scenarios} Based on participant feedback, \systemname is well-suited for three main scenarios. The first is ideation and reflective decision-making, such as brainstorming (P12) or simulating life choices (P4). The second is planning and comprehensive information organization, like creating travel lists (P9) or answering complex comparative questions (P6). The third is organizing information for academic and qualitative research, particularly for managing multiple layers of coding with unstructured data (P3). In summary, the tool is most beneficial for tasks requiring exploratory thinking, multi-layered analysis, and the structured output of initially ambiguous ideas.


\section{Discussion}

Our study demonstrates that \systemname{} offers distinct cognitive and interactive advantages over conversational LLM interfaces for speech-driven thought clarification. Rather than merely replicating known benefits of visual canvases~\cite{zhang2023visar, suh2023sensecape, pu2025ideasynth}, \systemname{} introduces a speech-first, spatially-aware sensemaking loop that uniquely combines fluid verbal expression and structured visual representation. Below, we interpret our findings through the lens of cognitive offloading, metacognitive scaffolding, and interaction modality trade-offs, and derive design implications that extend beyond prior work on AI-augmented thinking tools.

\subsection{Speech as a Primary Modality for Structured Thinking}
While prior systems have used speech for dictation or command, Orality treats speech as the raw material for continuous, structured ideation. This moves beyond Bolt's foundational ``Put-That-There'' \cite{bolt1980putthatthere} paradigm of speech as a control mechanism, and beyond tools like Rambler \cite{lin2024rambler} or MeetMap \cite{chen2025meetmap} that focus on producing a final artifact. Our observed user workflows—particularly the ``Master Commanders''—show that participants used speech not just to add content but to reconceptualize it through meta-instructions (e.g., ``\textit{merge these},'' ``\textit{organize this into these three parts}''...). This indicates that when coupled with a responsive visual representation, speech can become a powerful tool for iterative self-dialogue and conceptual restructuring. Our observed three distinct workflows reveal that speech can serve multiple roles: as a directive tool for AI-mediated restructuring, as a complement to manual spatial crafting of meaning, or as a low-barrier dumping mechanism. This plurality suggests that speech can serve as a flexible modality that adapts to users’ cognitive styles and task phases.

\subsection{The AI's Role: From Conversational Partner to Embedded Scaffold}
A key design challenge in AI-augmented thinking tools is balancing helpful automation with users' own thinking. Our work is contextualized in the evolving HCI discourse on tools for thought \cite{tankelevitchToolsThoughtResearch2025a}. Recent research has shown similar positive effects of AI asking questions \cite{10.1145/3544548.3580763} or guiding users based on their own way of thinking \cite{10.1145/3706598.3713295}. Orality's design offers a distinct approach by embedding AI assistance directly into the spatial representation of users' schematized thoughts rather than situating it in a separate conversational channel. The in-place ``Ask Me Questions'' feature, which users rated highly, acts as a cognitive probe within the user's own conceptual map. This transforms the AI's role from a conversational partner that can interrupt users' narratives (a noted drawback of the baseline) to a scaffold that extends and deepens users' thinking.

Existing AI-assisted canvas interface research has shown the benefits of spatial representation for divergent thinking, with enhanced monitoring and navigation support~\cite{zhang2023visar, suh2023sensecape, pu2025ideasynth}. Building on them, our work uniquely supports speech-first workflows with a real-time coupling between speech input and a malleable spatial workspace. It supports frequent shifts between the foraging (externalizing) and schematizing (structuring) stages in Pirolli and Card's sensemaking loop~\cite{pirolli2005sensemaking}. Our study identified how this approach can support thinking by facilitating divergent, in-depth thinking, active sensemaking, and targeted thought development.

Aligned with recent discussions advocating for ``generative friction''~\cite{Malaguti2025Investigating} or designing AI that rewards people to think~\cite{Rogers_2025}, our study showed that \systemname{} supported a more active sensemaking process by guiding users to think, even when the extracted topics were inaccurate (Section~\ref{active_sensemaking}), whereas the baseline directly provided answers for them. The fact that our system was recognized as better supporting thinking by 8 out of 12 participants showed promise in such approaches.

\subsection{Design implications}

The findings from our user study lead us to propose design implications for future tools that support thought clarification through speech and semantic interfaces.

\subsubsection{Building Adaptive Semantic Representation} 
While our semantic canvas was shown to be effective, participant feedback pointed to opportunities for more diverse and dynamic representations. Some users found the manual layout ``expensive'' (P10), while others desired more expressive spatial relationships, such as linking nodes across categories (P5) or creating different groupings from multiple perspectives (P7). This indicates that a simple free-form canvas with basic information extraction and schematization capabilities may not be sufficient for all tasks or users. Future systems should support multiple representational lenses (timeline, hierarchy, network, quadrant) that can be switched or merged based on task context. Users should be able to re-spatialize the same content under different semantic arrangements (e.g., by theme, by priority, by chronology), supporting multi-perspective sensemaking. Future systems should explore alternative and hybrid semantic spaces, such as:
\begin{itemize}
    \item Predefined Templates: Offering optional layouts (e.g., timelines, hierarchies, quadrant graphs) to provide scaffolding for common thought patterns.
    \item Semantic Zoom: Allowing users to collapse and expand node groups at different levels of abstraction.
    \item Multiple Views: Enabling the same set of thoughts to be visualized and interacted with through different structural lenses (e.g., by theme, by priority, by chronology) to support richer analysis.
\end{itemize}

\subsubsection{Designing Contextualized AI Suggestions}
Our study showed that AI-generated inspirational questions (the Ask Me Questions feature) were highly valued for stimulating new and deeper thoughts. Questions and prompts generated in situ and positioned as potential new nodes within underdeveloped topic areas help visually highlight gaps. This transforms AI suggestions from disruptive, out-of-context prompts into integrated, context-aware elements that flow naturally within the user's established spatial logic, reducing cognitive load and supporting a more continuous thought process. The Show Me Conflict feature was less well-used, which reveals design challenges and opportunities for detecting and showing complex semantic relationships. Future research could explore solutions to guide users in resolving conflicts. 
Moreover, while our system only explored spatial context, our participants wished our AI to be more proactive. Future research could open up the temporal dimension to provide more contextualized AI support for thinking.

\subsubsection{Improving Intent Recognition for Verbal Structuring} 
The power and frequent use of the Verbal Structuring feature, especially by the ``Master Commanders'' (P2, P5, P7, P8, P12), underscores a strong user need for high-level, language-based reorganization. Participants used commands not just for basic grouping but for advanced analytical tasks like synthesizing content to answer a specific question (P2:`` \textit{give me ideas on why I'm a good fit}'') or finding overlap between concepts (P12:`` \textit{list the overlapping parts}''). However, users also encountered limitations when the interpretation or execution didn't match their intent. This suggests a need for a more powerful and reliable method to improve this feature, perhaps through both interface and technical solutions, that allow users to more easily apply complex transformations—such as pivoting the entire canvas to a new perspective, merging large sections, or automatically applying a user-defined template structure. 

\subsubsection{Enhancing Metacognitive Support}
Part of the metacognitive demand comes from the potentially unpredictable outcome of Verbal Restructuring, given the freeform nature of user input. This can be potentially improved by providing shortcuts for common operations or user-defined reusable prompts. Our system was shown to be able to provide some metacognitive support, in ways to help clarify meanings, grasp the progress of thought development on topics, trace back thought evolution history, and generate perspectives. This is an important path forward as we enhance such support in future systems. For instance, we could design a more powerful Thought Evolution tool that allows thought inspection through user-defined dimensions and at multiple abstraction levels.
Instead of a single timeline, the system could allow users to create named checkpoints, branch off different thought trajectories, or filter the history to show only changes related to a specific topic. This would better support the metacognitive process of tracing the genesis of ideas, understanding how different decisions led to different outcomes, and revisiting abandoned paths without losing the current thread. The \textit{Export} with selected topics was used by several participants but lightly, due to the reading cost of the lengthy output. Perhaps future systems could improve this feature by making the output easier to review and find connections between multiple versions of exported output.

\subsection{Limitations and Future Work}


This study has several limitations that should be acknowledged. 

\subsubsection{Lab setting may limit realism and task authenticity} First, as a laboratory study, the experimental setting lacks ecological validity, and the thought clarification tasks that the participants came up with may have been slightly superficial. Future research should explore how to support thinking with such tools in more realistic tasks and real-world contexts.

\subsubsection{Variability in content creation limits comparison } 
While allowing participants to self-generate task topics was necessary to encourage a critical thinking process, 
it also introduced variability in task difficulty and goal orientation. Some topics aimed at a concrete solution, for example, preparing a travel plan, while others emphasized incremental progress in understanding a problem, e.g., a personal career path. This variation is difficult to control and may affect how participants perceive the tool's effectiveness in clarifying their thoughts.

\subsubsection{System output instability} Our prototype sometimes failed to reflect participants’ intended organization due to insufficient content context, unclear voice instruction, and LLMs' instability. 
Instances of misclassification, unintended merging or deletion of nodes, and incomplete conflict detection were observed. Participants frequently reported that the system failed to capture nuances of their intended organization, particularly in the ``Show Me Conflicts'' function, which sometimes failed to identify conflicts that were evident to the participants. Future work should improve the technical performance of AI support. 
\section{Conclusion}

This paper introduces \systemname, an AI-enhanced organic interface designed to support the use of speech and a semantic canvas for thought clarification. 
Addressing the challenges found in our formative study, \systemname offers a four-layer framework to (1) transform linear speech into a spatial, semantic representation; (2) provide a verbal structuring feature for flexible content reorganization; (3) stimulate deeper reflection and elaboration through on-demand in-place AI suggestions; and (4) visualize the iterative evolution of the thought process. Our user study showed that participants could successfully use \systemname for thought clarification and preferred it for supporting the thinking process over a prompt-enhanced LLM, demonstrating the potential of augmenting speech-driven thought clarification using a spatial AI interface. We envision a future where speech-first, canvas-based interfaces blend the fluidity of verbal expression with the structural power of visual semantics, ultimately augmenting our ability to think, create, and make sense of our most complex ideas.



\begin{acks}
This research is supported by a donation (project No. 9220140) from Huawei Technologies Co., Ltd., and the Google Faculty Research Award (project No. 9229068) from Google LLC.
We would like to thank Alvaro Cassinelli and Advait Sarkar for their inspiring discussions. Many thanks to our anonymous reviewers for their constructive and insightful feedback. 
We would like to add a note on the co-first authorship for transparency: Wengxi Li identified the theoretical framework, completed the system design and implementation. Jingze Tian led the design and conduct of the formative and evaluation studies. 
Both participated in the data analysis and the writing of the paper.

We acknowledge the use of ChatGPT in refining wordings and making it more concise throughout the paper with careful review of the researchers to make sure the original meanings remain unchanged. The authors take full responsibility for the correctness and authenticity of the content. 
\end{acks}

\bibliographystyle{ACM-Reference-Format}
\bibliography{references}

@String{Computing = "Computing" }

@String{Computer = "{IEEE} Computer" }

@String{Academic = "Academic Press" }

@String{Springer = "Springer-Verlag" }

@article{clark1998extended,
 ISSN = {00032638, 14678284},
 URL = {http://www.jstor.org/stable/3328150},
 author = {Andy Clark and David Chalmers},
 journal = {Analysis},
 number = {1},
 pages = {7--19},
 publisher = {[Analysis Committee, Oxford University Press]},
 title = {The Extended Mind},
 urldate = {2025-09-07},
 volume = {58},
 year = {1998}
}

@book{kahneman2011thinking,
  title={Thinking, fast and slow},
  author={Kahneman, Daniel},
  year={2011},
  publisher={Farrar, Straus and Giroux},
  address={New York, NY}
}

@article{reyna1995fuzzy,
  title={Fuzzy-trace theory: An interim synthesis},
  author={Reyna, Valerie F and Brainerd, Charles J},
  journal={Learning and individual Differences},
  volume={7},
  number={1},
  pages={1--75},
  year={1995},
  publisher={Elsevier}
}

@article{sweller1988cognitive,
  title={Cognitive load during problem solving: Effects on learning},
  author={Sweller, John},
  journal={Cognitive science},
  volume={12},
  number={2},
  pages={257--285},
  year={1988},
  publisher={Elsevier}
}

@inproceedings{pirolli2005sensemaking,
  title={The sensemaking process and leverage points for analyst technology as identified through cognitive task analysis},
  author={Pirolli, Peter and Card, Stuart},
  booktitle={Proceedings of international conference on intelligence analysis},
  volume={5},
  pages={2--4},
  year={2005},
  publisher={The MITRE Corporation},
  address={McLean, VA, USA}
}

@book{weick1995sensemaking,
  title={Sensemaking in organizations},
  author={Weick, Karl E.},
  volume={3},
  series={Foundations for Organizational Science},
  year={1995},
  publisher={Sage Publications},
  address={Thousand Oaks, CA}
}

@misc{zhang2024mindalogue,
      title={Mindalogue: LLM-Powered Nonlinear Interaction for Effective Learning and Task Exploration}, 
      author={Rui Zhang and Ziyao Zhang and Fengliang Zhu and Jiajie Zhou and Anyi Rao},
      year={2024},
      eprint={2410.10570},
      archivePrefix={arXiv},
      primaryClass={cs.HC},
      url={https://arxiv.org/abs/2410.10570}, 
}

@inproceedings{zhang2023visar,
author = {Zhang, Zheng and Gao, Jie and Dhaliwal, Ranjodh Singh and Li, Toby Jia-Jun},
title = {VISAR: A Human-AI Argumentative Writing Assistant with Visual Programming and Rapid Draft Prototyping},
year = {2023},
isbn = {9798400701320},
publisher = {Association for Computing Machinery},
address = {New York, NY, USA},
url = {https://doi.org/10.1145/3586183.3606800},
doi = {10.1145/3586183.3606800},
booktitle = {Proceedings of the 36th Annual ACM Symposium on User Interface Software and Technology},
articleno = {5},
numpages = {30},
location = {San Francisco, CA, USA},
series = {UIST '23}
}

@inproceedings{suh2024luminate,
author = {Suh, Sangho and Chen, Meng and Min, Bryan and Li, Toby Jia-Jun and Xia, Haijun},
title = {Luminate: Structured Generation and Exploration of Design Space with Large Language Models for Human-AI Co-Creation},
year = {2024},
isbn = {9798400703300},
publisher = {Association for Computing Machinery},
address = {New York, NY, USA},
url = {https://doi.org/10.1145/3613904.3642400},
doi = {10.1145/3613904.3642400},
booktitle = {Proceedings of the 2024 CHI Conference on Human Factors in Computing Systems},
articleno = {644},
numpages = {26},
location = {Honolulu, HI, USA},
series = {CHI '24}
}

@inproceedings{zhang2025ladica,
author = {Zhang, Zheng and Peng, Weirui and Chen, Xinyue and Cao, Luke and Li, Toby Jia-Jun},
title = {LADICA: A Large Shared Display Interface for Generative AI Cognitive Assistance in Co-located Team Collaboration},
year = {2025},
isbn = {9798400713941},
publisher = {Association for Computing Machinery},
address = {New York, NY, USA},
url = {https://doi.org/10.1145/3706598.3713289},
doi = {10.1145/3706598.3713289},
booktitle = {Proceedings of the 2025 CHI Conference on Human Factors in Computing Systems},
articleno = {147},
numpages = {22},
location = {
},
series = {CHI '25}
}

@inproceedings{lin2024rambler,
author = {Lin, Susan and Warner, Jeremy and Zamfirescu-Pereira, J.D. and Lee, Matthew G and Jain, Sauhard and Cai, Shanqing and Lertvittayakumjorn, Piyawat and Huang, Michael Xuelin and Zhai, Shumin and Hartmann, Bjoern and Liu, Can},
title = {Rambler: Supporting Writing With Speech via LLM-Assisted Gist Manipulation},
year = {2024},
isbn = {9798400703300},
publisher = {Association for Computing Machinery},
address = {New York, NY, USA},
url = {https://doi.org/10.1145/3613904.3642217},
doi = {10.1145/3613904.3642217},
booktitle = {Proceedings of the 2024 CHI Conference on Human Factors in Computing Systems},
articleno = {1043},
numpages = {19},
location = {Honolulu, HI, USA},
series = {CHI '24}
}

@inproceedings{lin2024videomap,
author = {Lin, David Chuan-En and Caba Heilbron, Fabian and Lee, Joon-Young and Wang, Oliver and Martelaro, Nikolas},
title = {VideoMap: Supporting Video Exploration, Brainstorming, and Prototyping in the Latent Space},
year = {2024},
isbn = {9798400704857},
publisher = {Association for Computing Machinery},
address = {New York, NY, USA},
url = {https://doi.org/10.1145/3635636.3656192},
doi = {10.1145/3635636.3656192},
booktitle = {Proceedings of the 16th Conference on Creativity \& Cognition},
pages = {311–327},
numpages = {17},
location = {Chicago, IL, USA},
series = {C\&C '24}
}

@inproceedings{suh2023sensecape,
author = {Suh, Sangho and Min, Bryan and Palani, Srishti and Xia, Haijun},
title = {Sensecape: Enabling Multilevel Exploration and Sensemaking with Large Language Models},
year = {2023},
isbn = {9798400701320},
publisher = {Association for Computing Machinery},
address = {New York, NY, USA},
url = {https://doi.org/10.1145/3586183.3606756},
doi = {10.1145/3586183.3606756},
booktitle = {Proceedings of the 36th Annual ACM Symposium on User Interface Software and Technology},
articleno = {1},
numpages = {18},
location = {San Francisco, CA, USA},
series = {UIST '23}
}

@inproceedings{zhang2021conceptscope,
author = {Zhang, Xiaoyu and Chandrasegaran, Senthil and Ma, Kwan-Liu},
title = {ConceptScope: Organizing and Visualizing Knowledge in Documents based on Domain Ontology},
year = {2021},
isbn = {9781450380966},
publisher = {Association for Computing Machinery},
address = {New York, NY, USA},
url = {https://doi.org/10.1145/3411764.3445396},
doi = {10.1145/3411764.3445396},
booktitle = {Proceedings of the 2021 CHI Conference on Human Factors in Computing Systems},
articleno = {19},
numpages = {13},
location = {Yokohama, Japan},
series = {CHI '21}
}

@inproceedings{zhang2023concepteva,
author = {Zhang, Xiaoyu and Li, Jianping and Chi, Po-Wei and Chandrasegaran, Senthil and Ma, Kwan-Liu},
title = {ConceptEVA: Concept-Based Interactive Exploration and Customization of Document Summaries},
year = {2023},
isbn = {9781450394215},
publisher = {Association for Computing Machinery},
address = {New York, NY, USA},
url = {https://doi.org/10.1145/3544548.3581260},
doi = {10.1145/3544548.3581260},
booktitle = {Proceedings of the 2023 CHI Conference on Human Factors in Computing Systems},
articleno = {204},
numpages = {16},
location = {Hamburg, Germany},
series = {CHI '23}
}

@inproceedings{chung2024patchview,
author = {Chung, John Joon Young and Kreminski, Max},
title = {Patchview: LLM-powered Worldbuilding with Generative Dust and Magnet Visualization},
year = {2024},
isbn = {9798400706288},
publisher = {Association for Computing Machinery},
address = {New York, NY, USA},
url = {https://doi.org/10.1145/3654777.3676352},
doi = {10.1145/3654777.3676352},
booktitle = {Proceedings of the 37th Annual ACM Symposium on User Interface Software and Technology},
articleno = {77},
numpages = {19},
location = {Pittsburgh, PA, USA},
series = {UIST '24}
}

@inproceedings{chandrasegaran2019talktraces,
author = {Chandrasegaran, Senthil and Bryan, Chris and Shidara, Hidekazu and Chuang, Tung-Yen and Ma, Kwan-Liu},
title = {TalkTraces: Real-Time Capture and Visualization of Verbal Content in Meetings},
year = {2019},
isbn = {9781450359702},
publisher = {Association for Computing Machinery},
address = {New York, NY, USA},
url = {https://doi.org/10.1145/3290605.3300807},
doi = {10.1145/3290605.3300807},
booktitle = {Proceedings of the 2019 CHI Conference on Human Factors in Computing Systems},
pages = {1–14},
numpages = {14},
location = {Glasgow, Scotland Uk},
series = {CHI '19}
}

@inproceedings{xing2025immersed,
author = {Xing, Yunhao and Ban, Jerrick and Hubbard, Timothy D and Villano, Michael and G\'{o}mez-Zar\'{a}, Diego},
title = {Immersed in my Ideas: Using Virtual Reality and LLMs to Visualize Users' Ideas and Thoughts},
year = {2025},
isbn = {9798400714092},
publisher = {Association for Computing Machinery},
address = {New York, NY, USA},
url = {https://doi.org/10.1145/3708557.3716330},
doi = {10.1145/3708557.3716330},
booktitle = {Companion Proceedings of the 30th International Conference on Intelligent User Interfaces},
pages = {60–65},
numpages = {6},
location = {
},
series = {IUI '25 Companion}
}

@inproceedings{jiang2023graphologue,
author = {Jiang, Peiling and Rayan, Jude and Dow, Steven P. and Xia, Haijun},
title = {Graphologue: Exploring Large Language Model Responses with Interactive Diagrams},
year = {2023},
isbn = {9798400701320},
publisher = {Association for Computing Machinery},
address = {New York, NY, USA},
url = {https://doi.org/10.1145/3586183.3606737},
doi = {10.1145/3586183.3606737},
booktitle = {Proceedings of the 36th Annual ACM Symposium on User Interface Software and Technology},
articleno = {3},
numpages = {20},
location = {San Francisco, CA, USA},
series = {UIST '23}
}

@article{flower1981writing,
 ISSN = {0010096X},
 URL = {http://www.jstor.org/stable/356600},
 author = {Linda Flower and John R. Hayes},
 journal = {College Composition and Communication},
 number = {4},
 pages = {365--387},
 publisher = {National Council of Teachers of English},
 title = {A Cognitive Process Theory of Writing},
 urldate = {2025-09-07},
 volume = {32},
 year = {1981}
}

@inproceedings{lee2024writing,
author = {Lee, Mina and Gero, Katy Ilonka and Chung, John Joon Young and Shum, Simon Buckingham and Raheja, Vipul and Shen, Hua and Venugopalan, Subhashini and Wambsganss, Thiemo and Zhou, David and Alghamdi, Emad A. and August, Tal and Bhat, Avinash and Choksi, Madiha Zahrah and Dutta, Senjuti and Guo, Jin L.C. and Hoque, Md Naimul and Kim, Yewon and Knight, Simon and Neshaei, Seyed Parsa and Shibani, Antonette and Shrivastava, Disha and Shroff, Lila and Sergeyuk, Agnia and Stark, Jessi and Sterman, Sarah and Wang, Sitong and Bosselut, Antoine and Buschek, Daniel and Chang, Joseph Chee and Chen, Sherol and Kreminski, Max and Park, Joonsuk and Pea, Roy and Rho, Eugenia Ha Rim and Shen, Zejiang and Siangliulue, Pao},
title = {A Design Space for Intelligent and Interactive Writing Assistants},
year = {2024},
isbn = {9798400703300},
publisher = {Association for Computing Machinery},
address = {New York, NY, USA},
url = {https://doi.org/10.1145/3613904.3642697},
doi = {10.1145/3613904.3642697},
booktitle = {Proceedings of the 2024 CHI Conference on Human Factors in Computing Systems},
articleno = {1054},
numpages = {35},
location = {Honolulu, HI, USA},
series = {CHI '24}
}

@article{kirsh2010thinking,
  title={Thinking with external representations},
  author={Kirsh, David},
  journal={AI \& society},
  volume={25},
  number={4},
  pages={441--454},
  year={2010},
  publisher={Springer}
}

@article{ericsson1980verbal,
  title={Verbal reports as data.},
  author={Ericsson, K Anders and Simon, Herbert A},
  journal={Psychological review},
  volume={87},
  number={3},
  pages={215},
  year={1980},
  publisher={American Psychological Association}
}

@article{chi1994eliciting,
  title={Eliciting self-explanations improves understanding},
  author={Chi, Michelene TH and De Leeuw, Nicholas and Chiu, Mei-Hung and LaVancher, Christian},
  journal={Cognitive science},
  volume={18},
  number={3},
  pages={439--477},
  year={1994},
  publisher={Elsevier}
}

@article{sripada2020structure,
  title={Structure in the stream of consciousness: Evidence from a verbalized thought protocol and automated text analytic methods},
  author={Sripada, Chandra and Taxali, Aman},
  journal={Consciousness and Cognition},
  volume={85},
  pages={103007},
  year={2020},
  publisher={Elsevier}
}

@inproceedings{lev2024metacognitive,
author = {Tankelevitch, Lev and Kewenig, Viktor and Simkute, Auste and Scott, Ava Elizabeth and Sarkar, Advait and Sellen, Abigail and Rintel, Sean},
title = {The Metacognitive Demands and Opportunities of Generative AI},
year = {2024},
isbn = {9798400703300},
publisher = {Association for Computing Machinery},
address = {New York, NY, USA},
url = {https://doi.org/10.1145/3613904.3642902},
doi = {10.1145/3613904.3642902},
booktitle = {Proceedings of the 2024 CHI Conference on Human Factors in Computing Systems},
articleno = {680},
numpages = {24},
location = {Honolulu, HI, USA},
series = {CHI '24}
}

@article{gerlich2025ai,
  title={AI tools in society: Impacts on cognitive offloading and the future of critical thinking},
  author={Gerlich, Michael},
  journal={Societies},
  doi = {10.3390/soc15010006},
  volume={15},
  number={1},
  pages={6},
  year={2025},
  publisher={Multidisciplinary Digital Publishing Institute}
}

@article{lin2024decision,
    title = "Decision-Oriented Dialogue for Human-{AI} Collaboration",
    author = "Lin, Jessy  and
      Tomlin, Nicholas  and
      Andreas, Jacob  and
      Eisner, Jason",
    journal = "Transactions of the Association for Computational Linguistics",
    volume = "12",
    year = "2024",
    address = "Cambridge, MA",
    publisher = "MIT Press",
    url = "https://aclanthology.org/2024.tacl-1.50/",
    doi = "10.1162/tacl_a_00679",
    pages = "892--911"
}

@article{Daryanto2025conversate,
author = {Daryanto, Taufiq and Ding, Xiaohan and Wilhelm, Lance T. and Stil, Sophia and Knutsen, Kirk McInnis and Rho, Eugenia H.},
title = {Conversate: Supporting Reflective Learning in Interview Practice Through Interactive Simulation and Dialogic Feedback},
year = {2025},
issue_date = {January 2025},
publisher = {Association for Computing Machinery},
address = {New York, NY, USA},
volume = {9},
number = {1},
url = {https://doi.org/10.1145/3701188},
doi = {10.1145/3701188},
journal = {Proc. ACM Hum.-Comput. Interact.},
month = jan,
articleno = {GROUP9},
numpages = {32}
}

@article{homer2025comparative,
  title={Comparative analysis of concept mapping: human participants vs. ChatGPT},
  author={Homer, Stephen T},
  journal={Quality \& Quantity},
  pages={4873--4892},
  year={2025},
  doi={10.1007/s11135-025-02211-w},
  publisher={Springer},
  volume={59}
}

@article{schicchi2025closer,
  title     = {A closer look at ChatGPT’s role in concept map generation for education},
  author    = {Schicchi, Daniele and Limongelli, Carla and Monteleone, Vito and Taibi, Davide},
  journal   = {Interactive Learning Environments},
  volume    = {34},
  number    = {1},
  pages     = {276--296},
  year      = {2026},
  doi       = {10.1080/10494820.2025.2497110},
  publisher = {Taylor \& Francis}
}

@article{ruan2018comparing,
author = {Ruan, Sherry and Wobbrock, Jacob O. and Liou, Kenny and Ng, Andrew and Landay, James A.},
title = {Comparing Speech and Keyboard Text Entry for Short Messages in Two Languages on Touchscreen Phones},
year = {2018},
issue_date = {December 2017},
publisher = {Association for Computing Machinery},
address = {New York, NY, USA},
volume = {1},
number = {4},
url = {https://doi.org/10.1145/3161187},
doi = {10.1145/3161187},
journal = {Proc. ACM Interact. Mob. Wearable Ubiquitous Technol.},
month = jan,
articleno = {159},
numpages = {23}
}

@article{arons1997speechskimmer,
author = {Arons, Barry},
title = {SpeechSkimmer: a system for interactively skimming recorded speech},
year = {1997},
issue_date = {March 1997},
publisher = {Association for Computing Machinery},
address = {New York, NY, USA},
volume = {4},
number = {1},
issn = {1073-0516},
url = {https://doi.org/10.1145/244754.244758},
doi = {10.1145/244754.244758},
journal = {ACM Trans. Comput.-Hum. Interact.},
month = mar,
pages = {3–38},
numpages = {36}
}

@ArtifactSoftware{Zoom,
    title = {Zoom},
    author = {{Zoom Core Team}},
    organization = {Zoom},
    address = {California, United States},
    year = {2025},
    url = {https://www.zoom.com/},
}

@article{evans2008dual,
  title={Dual-processing accounts of reasoning, judgment, and social cognition},
  author={Evans, Jonathan St BT},
  journal={Annu. Rev. Psychol.},
  volume={59},
  number={1},
  pages={255--278},
  year={2008},
  publisher={Annual Reviews}
}

@inproceedings{huang2024conversation,
author = {Huang, Shih-Hong and Lin, Ya-Fang and He, Zeyu and Huang, Chieh-Yang and Huang, Ting-Hao Kenneth},
title = {How Does Conversation Length Impact User’s Satisfaction? A Case Study of Length-Controlled Conversations with LLM-Powered Chatbots},
year = {2024},
isbn = {9798400703317},
publisher = {Association for Computing Machinery},
address = {New York, NY, USA},
url = {https://doi.org/10.1145/3613905.3650823},
doi = {10.1145/3613905.3650823},
booktitle = {Extended Abstracts of the CHI Conference on Human Factors in Computing Systems},
articleno = {188},
numpages = {13},
location = {Honolulu, HI, USA},
series = {CHI EA '24}
}

@article{flavell1979metacognition,
  title={Metacognition and cognitive monitoring: A new area of cognitive--developmental inquiry.},
  author={Flavell, John H},
  journal={American psychologist},
  doi={10.1037/0003-066X.34.10.906},
  volume={34},
  number={10},
  pages={906},
  year={1979},
  publisher={American Psychological Association}
}

@incollection{nelson1990metamemory,
  title={Metamemory: A theoretical framework and new findings},
  author={Nelson, Thomas O},
  booktitle={Psychology of learning and motivation},
  volume={26},
  pages={125--173},
  year={1990},
  publisher={Elsevier},
  doi={10.1016/S0079-7421(08)60053-5},
  address = {San Diego, CA}
}

@misc{tankelevitch2025understanding,
      title={Understanding, Protecting, and Augmenting Human Cognition with Generative AI: A Synthesis of the CHI 2025 Tools for Thought Workshop}, 
      author={Lev Tankelevitch and Elena L. Glassman and Jessica He and Aniket Kittur and Mina Lee and Srishti Palani and Advait Sarkar and Gonzalo Ramos and Yvonne Rogers and Hari Subramonyam},
      year={2025},
      eprint={2508.21036},
      archivePrefix={arXiv},
      primaryClass={cs.HC},
      url={https://arxiv.org/abs/2508.21036}, 
}

@misc{singh2025enhancing,
      title={Enhancing Critical Thinking in Generative AI Search with Metacognitive Prompts}, 
      author={Anjali Singh and Zhitong Guan and Soo Young Rieh},
      year={2025},
      eprint={2505.24014},
      archivePrefix={arXiv},
      primaryClass={cs.HC},
      url={https://arxiv.org/abs/2505.24014}, 
}

@article{wang2023scaffolding,
author = {Wang, Cui-Yu and Gao, Bao-Lian and Chen, Shu-Jie},
title = {The effects of metacognitive scaffolding of project-based learning environments on students’ metacognitive ability and computational thinking},
year = {2023},
issue_date = {Apr 2024},
publisher = {Kluwer Academic Publishers},
address = {USA},
volume = {29},
number = {5},
issn = {1360-2357},
url = {https://doi.org/10.1007/s10639-023-12022-x},
doi = {10.1007/s10639-023-12022-x},
journal = {Education and Information Technologies},
month = jul,
pages = {5485–5508},
numpages = {24}
}

@misc{yatani2024aiextraherics,
      title={AI as Extraherics: Fostering Higher-order Thinking Skills in Human-AI Interaction}, 
      author={Koji Yatani and Zefan Sramek and Chi-Lan Yang},
      year={2024},
      eprint={2409.09218},
      archivePrefix={arXiv},
      primaryClass={cs.HC},
      url={https://arxiv.org/abs/2409.09218}, 
}

@misc{liao2023aitransparency,
      title={AI Transparency in the Age of LLMs: A Human-Centered Research Roadmap}, 
      author={Q. Vera Liao and Jennifer Wortman Vaughan},
      year={2023},
      eprint={2306.01941},
      archivePrefix={arXiv},
      primaryClass={cs.HC},
      url={https://arxiv.org/abs/2306.01941}, 
}

@techreport{novak2008theory,
  title       = {The Theory Underlying Concept Maps and How to Construct and Use Them},
  author      = {Novak, Joseph D. and Ca{\~n}as, Alberto J.},
  institution = {Florida Institute for Human and Machine Cognition},
  year        = {2008},
  number      = {IHMC CmapTools 2006-01 Rev 01-2008},
  address     = {Pensacola, FL},
  pages       = {1--31},
  numpages    = {31},
  url         = {http://cmap.ihmc.us/Publications/ResearchPapers/TheoryUnderlyingConceptMaps.pdf}
}

@article{asthana2025summaries,
author = {Asthana, Sumit and Hilleli, Sagi and He, Pengcheng and Halfaker, Aaron},
title = {Summaries, Highlights, and Action Items: Design, Implementation and Evaluation of an LLM-powered Meeting Recap System},
year = {2025},
issue_date = {May 2025},
publisher = {Association for Computing Machinery},
address = {New York, NY, USA},
volume = {9},
number = {2},
url = {https://doi.org/10.1145/3711074},
doi = {10.1145/3711074},
journal = {Proc. ACM Hum.-Comput. Interact.},
month = may,
articleno = {CSCW176},
numpages = {29}
}

@article{chen2025meetmap,
author = {Chen, Xinyue and Yap, Nathan and Lu, Xinyi and Gunal, Aylin and Wang, Xu},
title = {MeetMap: Real-Time Collaborative Dialogue Mapping with LLMs in Online Meetings},
year = {2025},
issue_date = {May 2025},
publisher = {Association for Computing Machinery},
address = {New York, NY, USA},
volume = {9},
number = {2},
url = {https://doi.org/10.1145/3711030},
doi = {10.1145/3711030},
journal = {Proc. ACM Hum.-Comput. Interact.},
month = may,
articleno = {CSCW132},
numpages = {35}
}

@inproceedings{bolt1980putthatthere,
author = {Bolt, Richard A.},
title = {“Put-that-there”: Voice and gesture at the graphics interface},
year = {1980},
isbn = {0897910214},
publisher = {Association for Computing Machinery},
address = {New York, NY, USA},
url = {https://doi.org/10.1145/800250.807503},
doi = {10.1145/800250.807503},
booktitle = {Proceedings of the 7th Annual Conference on Computer Graphics and Interactive Techniques},
pages = {262–270},
numpages = {9},
location = {Seattle, Washington, USA},
series = {SIGGRAPH '80}
}

@inproceedings{pu2025ideasynth,
author = {Pu, Kevin and Feng, K. J. Kevin and Grossman, Tovi and Hope, Tom and Dalvi Mishra, Bhavana and Latzke, Matt and Bragg, Jonathan and Chang, Joseph Chee and Siangliulue, Pao},
title = {IdeaSynth: Iterative Research Idea Development Through Evolving and Composing Idea Facets with Literature-Grounded Feedback},
year = {2025},
isbn = {9798400713941},
publisher = {Association for Computing Machinery},
address = {New York, NY, USA},
url = {https://doi.org/10.1145/3706598.3714057},
doi = {10.1145/3706598.3714057},
booktitle = {Proceedings of the 2025 CHI Conference on Human Factors in Computing Systems},
articleno = {145},
numpages = {31},
location = {
},
series = {CHI '25}
}

@inproceedings{Malaguti2025Investigating,
  author = {Malaguti, Pauline and Karran, Alexander J. and Le, Di and Mortin, Hayley and Coursaris, Constantinos K. and S{\'e}n{\'e}cal, Sylvain and L{\'e}ger, Pierre-Majorique},
  title = {Investigating Interaction Friction in Generative AI: Improving User Experience and Decision-Making},
  booktitle = {SIGHCI 2024 Proceedings},
  year = {2025},
  pages = {26},
  publisher = {Association for Information Systems (AIS)},
  address = {Bangkok, Thailand},
  url = {https://aisel.aisnet.org/sighci2024/26}
}

@article{Rogers_2025,
   title={Why It Is Worth Making An Effort When Learning With GEN AI},
   volume={06},
   ISSN={2693-2555},
   url={http://dx.doi.org/10.56734/ijahss.v6nSa1},
   DOI={10.56734/ijahss.v6nsa1},
   number={Special Issue},
   journal={International Journal of Arts , Humanities \&amp; Social Science},
   publisher={Institute for Promoting Research and Policy Development},
   author={Rogers, Yvonne},
   year={2025},
   month=aug, pages={1–6} }

@inproceedings{tankelevitchToolsThoughtResearch2025a,
author = {Tankelevitch, Lev and Glassman, Elena L. and He, Jessica and Kazemitabaar, Majeed and Kittur, Aniket and Lee, Mina and Palani, Srishti and Sarkar, Advait and Ramos, Gonzalo and Rogers, Yvonne and Subramonyam, Hari},
title = {Tools for Thought: Research and Design for Understanding, Protecting, and Augmenting Human Cognition with Generative AI},
year = {2025},
isbn = {9798400713958},
publisher = {Association for Computing Machinery},
address = {New York, NY, USA},
url = {https://doi.org/10.1145/3706599.3706745},
doi = {10.1145/3706599.3706745},
booktitle = {Proceedings of the Extended Abstracts of the CHI Conference on Human Factors in Computing Systems},
articleno = {804},
numpages = {8},
location = {
},
series = {CHI EA '25}
}

@inproceedings{10.1145/3544548.3580763,
author = {Wagener, Nadine and Reicherts, Leon and Zargham, Nima and Bart\l{}omiejczyk, Natalia and Scott, Ava Elizabeth and Wang, Katherine and Bentvelzen, Marit and Stefanidi, Evropi and Mildner, Thomas and Rogers, Yvonne and Niess, Jasmin},
title = {SelVReflect: A Guided VR Experience Fostering Reflection on Personal Challenges},
year = {2023},
isbn = {9781450394215},
publisher = {Association for Computing Machinery},
address = {New York, NY, USA},
url = {https://doi.org/10.1145/3544548.3580763},
doi = {10.1145/3544548.3580763},
booktitle = {Proceedings of the 2023 CHI Conference on Human Factors in Computing Systems},
articleno = {323},
numpages = {17},
location = {Hamburg, Germany},
series = {CHI '23}
}

@inproceedings{10.1145/3706598.3713295,
author = {Reicherts, Leon and Zhang, Zelun Tony and von Oswald, Elisabeth and Liu, Yuanting and Rogers, Yvonne and Hassib, Mariam},
title = {AI, Help Me Think—but for Myself: Assisting People in Complex Decision-Making by Providing Different Kinds of Cognitive Support},
year = {2025},
isbn = {9798400713941},
publisher = {Association for Computing Machinery},
address = {New York, NY, USA},
url = {https://doi.org/10.1145/3706598.3713295},
doi = {10.1145/3706598.3713295},
booktitle = {Proceedings of the 2025 CHI Conference on Human Factors in Computing Systems},
articleno = {255},
numpages = {19},
location = {
},
series = {CHI '25}
}

@book{schon1983reflective,
  title = {The Reflective Practitioner: How Professionals Think in Action},
  author = {Sch{\"o}n, Donald A.},
  year = {1983},
  publisher = {Basic Books},
  address = {New York},
  isbn = {0465068782}
}

@article{allen1999mixinitiative,
  author={Allen, J.E. and Guinn, C.I. and Horvtz, E.},
  journal={IEEE Intelligent Systems and their Applications}, 
  title={Mixed-initiative interaction}, 
  year={1999},
  volume={14},
  number={5},
  pages={14-23},
  doi={10.1109/5254.796083}}

@article{scaife1996external,
  title     = {External cognition: how do graphical representations work?},
  author    = {Scaife, Mike and Rogers, Yvonne},
  journal   = {International Journal of Human-Computer Studies},
  volume    = {45},
  number    = {2},
  pages     = {185--213},
  year      = {1996},
  publisher = {Elsevier},
  doi       = {10.1006/ijhc.1996.0048}
}

\clearpage
\appendix
\section{The Pirolli and Card Model}
\label{Pirolli and Card Model}

\begin{figure}[ht]
  \centering
  \includegraphics[width=\linewidth]{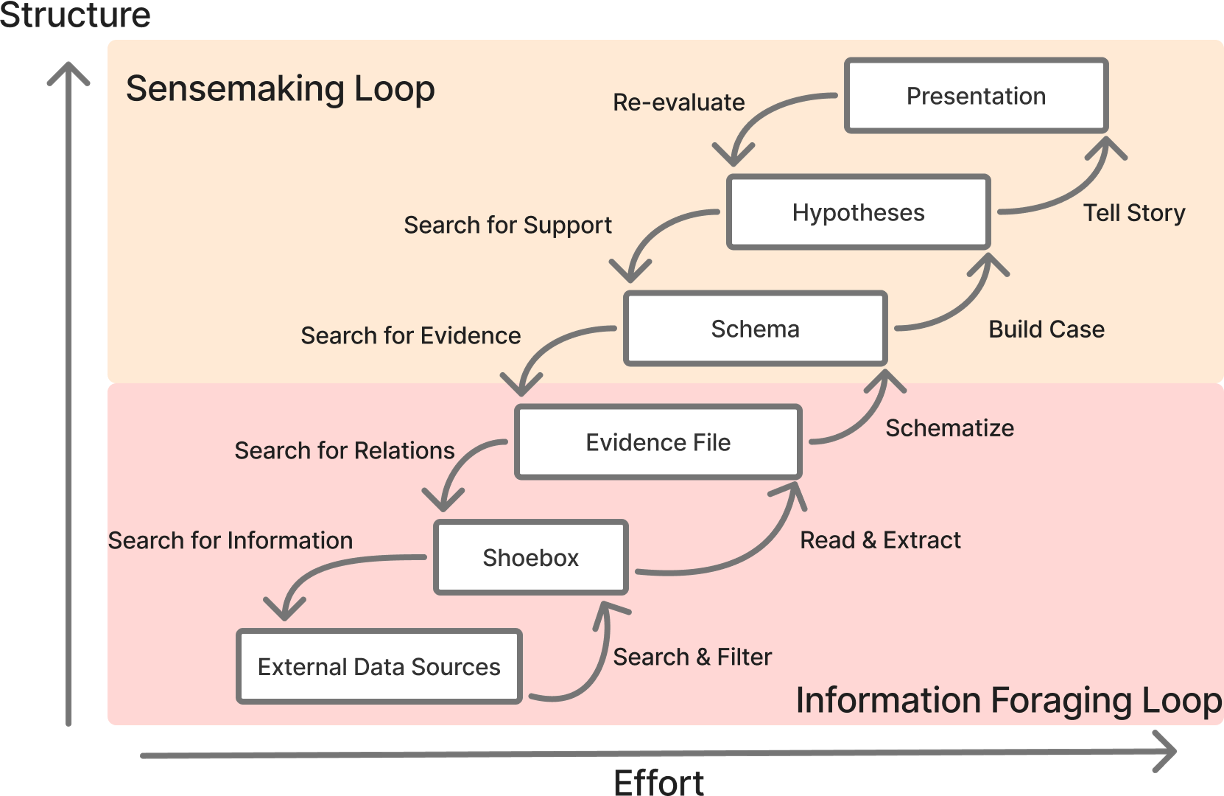}
  \caption{The original model by Pirolli and Card~\cite{pirolli2005sensemaking}.}
  \Description{A framework diagram illustrating the relationship between "Sensemaking Loop" and "Information Foraging Loop" processes, progressing from external data sources through various analytical stages to final presentation with increasing structure and effort.}
  \label{fig:originalmodel}
\end{figure}

\section{An example transcript of verbalized thoughts from the formative study}
\label{Formative Example}

P2 spoke about a research project that uses an Augmented Reality (AR) music game to gamify housework. The monologue captures the unstructured thoughts going through the project's several stages. 

``\textit{Um, recently I read a few papers, mainly about, uh, the challenges... then I found that, well, they aren't really able to enjoy and focus... so I'm considering... using games, especially augmented reality games, to help users... right now, my supervisor seems to think the direction is quite interesting, but the main problem for me is still, uh, the research questions... uh, first of all, well, how should I put it?... there might be three research questions... uh, and there's another research question, what was it again? I kinda forgot... uh, and now my idea is... I need to create a prototype first... I found that music games, uh, are a pretty good choice... it doesn't have any violent elements... it's simply following the music to, uh, tap to the beat... uh, and also... music games can get the user's body and hands moving... I'm considering designing, uh, an AR scenario where users play a game while cleaning windows... Well, anyway, this proposal, it's a course assignment... I need to plan its timeline... I think now I can first write down, uh, how I want to design this game... beat points... energy bar... uh, I think right before the moment they need to be clicked, there should be an animation... Hmm, I think I can consider the beat points this way... I need to control the difficulty... I can't make it too hard. Otherwise, the user might get too tired... uh, so I read before that difficulty control... Hmm... finally, how should I run the experiment? I need users to fill out two questionnaires, one about focus, one about enjoyment... objective data is important... uh, how clean the window is before and after, but that's hard to quantify... the time the user takes to complete the task... Feels quite tight. But we have a few people in the team now, maybe we work together... Okay, I think. That's probably about it.}''

\section{Extra details of the evaluation study}

\subsection{An example conversation snippet with the Baseline}
\label{Baseline example}

\begin{figure}[ht]
  \includegraphics[width=\columnwidth]{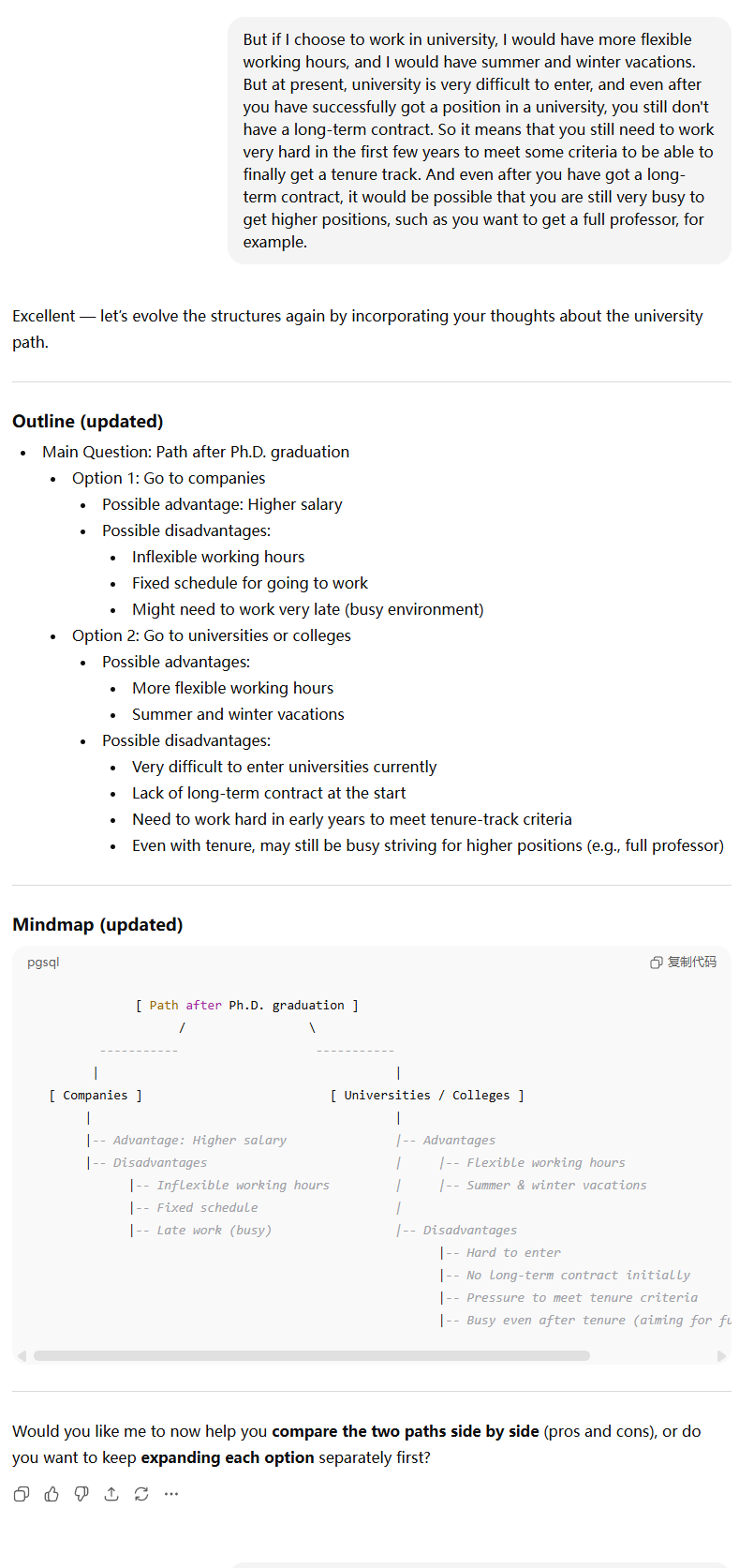}
  \caption{An example excerpt from a participant's task trials with baseline.}
  \Description{The image presents an excerpt from a participant's task in a thought clarification trial. It includes an outline and a mind map to help visualize two career paths after a Ph.D.: working at a company or in academia. The outline lists the pros and cons of each option, while the mind map visually organizes these options in a structured format. The facilitator also asks the participant whether to compare the paths side by side or continue expanding on each option separately.}
  \label{fig:baseline conversation}
\end{figure}

\clearpage
\onecolumn

\subsection{Participant Demographics}

\begin{table*}[h]
  \centering
  \caption{Demography and Dictation and GenAI usage of participants}
  \label{demographic data}
  \resizebox{0.6\textwidth}{!}{%
  \begin{tabular}{l|l|l|l|l|l}
  \hline
  \textbf{Participants} & \textbf{Attendence} & \textbf{Degree} & \textbf{Domain} & \textbf{Dictation Usage} & \textbf{GenAI Usage} \\ \hline
  P1 & Online & Master & HCI & yearly & Expert \\
  P2 & In-person & Undergrad & Design & monthly & Expert \\
  P3 & Online & Master & Health Info. & monthly & Expert \\
  P4 & In-person & Master & HCI & daily & Regular User \\
  P5 & In-person & Master & HCI & weekly & Casual User \\
  P6 & In-person & Master & HCI & monthly & Regular User \\
  P7 & In-person & Master & Comp. Sci. & weekly & Regular User \\
  P8 & In-person & Master & HCI & daily & Expert \\
  P9 & Online & Master & Ind. Design & daily & Casual User \\
  P10 & Online & Master & Comp. Sci. & yearly & Expert \\
  P11 & Online & Master & HCI & yearly & Regular User \\
  P12 & In-person & Undergrad & Game Design & never & Regular User \\ \hline
  \end{tabular}%
  }
  \Description{A demographic summary table showing characteristics of 12 study participants (P1-P12), including their study attendance mode (online vs in-person), degree level (mostly Master's with two undergraduates), academic domains (predominantly HCI, with representation from design, computer science, health informatics, and game design), frequency of dictation tool usage (ranging from never to daily), and self-reported GenAI expertise levels (Expert, Regular User, or Casual User).}
\end{table*}

\subsection{Participant Topics for the Evaluation Study} \label{sec:topic for study}

The topics and detailed information for each participant in the evaluation study are listed in Table \ref{tab:topics-orality} and \ref{tab:topics-baseline}. To protect participant privacy, specific identifying details such as exact locations, unique research project descriptions, and personal business ideas have been anonymized or generalized.

\begin{table*}[h]
\centering
\caption{Participants' topics in Orality Condition}
\Description{This table lists the topics selected by participants under the Orality Condition, categorized by the type of cognitive task. Each participant (P1 to P12) is assigned a specific subject, covering Strategic Planning (e.g., career roadmaps), Content Structuring (e.g., self-introductions), Reflective Sensemaking (e.g., discussing project challenges), Comparative Decision Making (e.g., job vs. PhD), and Problem Solving. These topics reflect a blend of career aspirations, academic progression, and future life planning.}
\label{tab:topics-orality}
\small
\resizebox{\textwidth}{!}{%
\begin{tabular}{l|p{0.6\textwidth}|l}
\hline
\textbf{Participant} & \textbf{Orality Condition Topic} & \textbf{Task Category} \\ \hline
P1 & How to become a software architect & Strategy Making \\
P2 & Preparing a self-introduction for a UX researcher internship interview & Strategy Making \\
P3 & Talking about an ongoing project and the current challenges & Project Planning \\
P4 & Career development choices after completing a PhD abroad & Comparative Decision Making \\
P5 & Planning the selection of a classification framework for a research project & Project Planning \\
P6 & Comparing quality of life and career prospects between two major cities & Comparative Decision Making \\
P7 & Developing a poster-generation application and writing a report & Project Planning \\
P8 & Summarizing three projects and planning for future work & Project Planning \\
P9 & Talking about an ongoing project in an industry setting & Project Planning \\
P10 & Weighing the choice between looking for a job or pursuing a PhD & Comparative Decision Making \\
P11 & Thinking about the study and life plan for the upcoming semester & Strategy Making \\
P12 & Personal growth and 5-year plan & Strategy Making \\ \hline
\end{tabular}%
}
\end{table*}

\begin{table*}[h]
\centering
\caption{Participants' topics in Baseline Condition}
\Description{This table outlines the topics chosen by participants under the Baseline Condition, categorized by task type. The topics include Comparative Decision Making (e.g., choosing cities or career paths), Problem Solving (e.g., saving money, sleep issues), Strategic Planning (e.g., research roadmaps, trip itineraries), and Reflective Sensemaking. The baseline topics focus on decision-making, planning, and problem-solving in personal and professional contexts.}
\label{tab:topics-baseline}
\small
\resizebox{\textwidth}{!}{%
\begin{tabular}{l|p{0.6\textwidth}|l}
\hline
\textbf{Participant} & \textbf{Baseline Condition Topic} & \textbf{Task Category} \\ \hline
P1 & Personal Development: Choosing between City A or City B & Comparative Decision Making \\
P2 & Planning how to save money as an international student & Strategy Making \\
P3 & Finding the reason behind having trouble falling asleep and how to solve it & Strategy Making 
\\
P4 & Outlining the research roadmap for a specified HCI topic & Project Planning \\
P5 & Planning a multi-day group trip itinerary, covering logistics & Strategy Making \\
P6 & Deciding between an academic career or an industry position after a PhD & Comparative Decision Making \\
P7 & Organizing a debate on educational methods for teenagers & Strategy Making \\
P8 & Deciding between seeking employment or pursuing a postdoctoral position & Comparative Decision Making \\
P9 & Talking about an ongoing academic research project & Project Planning \\
P10 & Describing an ongoing project and outlining its next step & Project Planning \\
P11 & Describing an ongoing project and scheduling tasks for the next two weeks & Project Planning \\
P12 & Developing a moneymaking plan invites others to become business partners & Strategy Making \\ \hline
\end{tabular}%
}
\end{table*}

\begin{figure*}
  \centering
  \includegraphics[width=\textwidth]{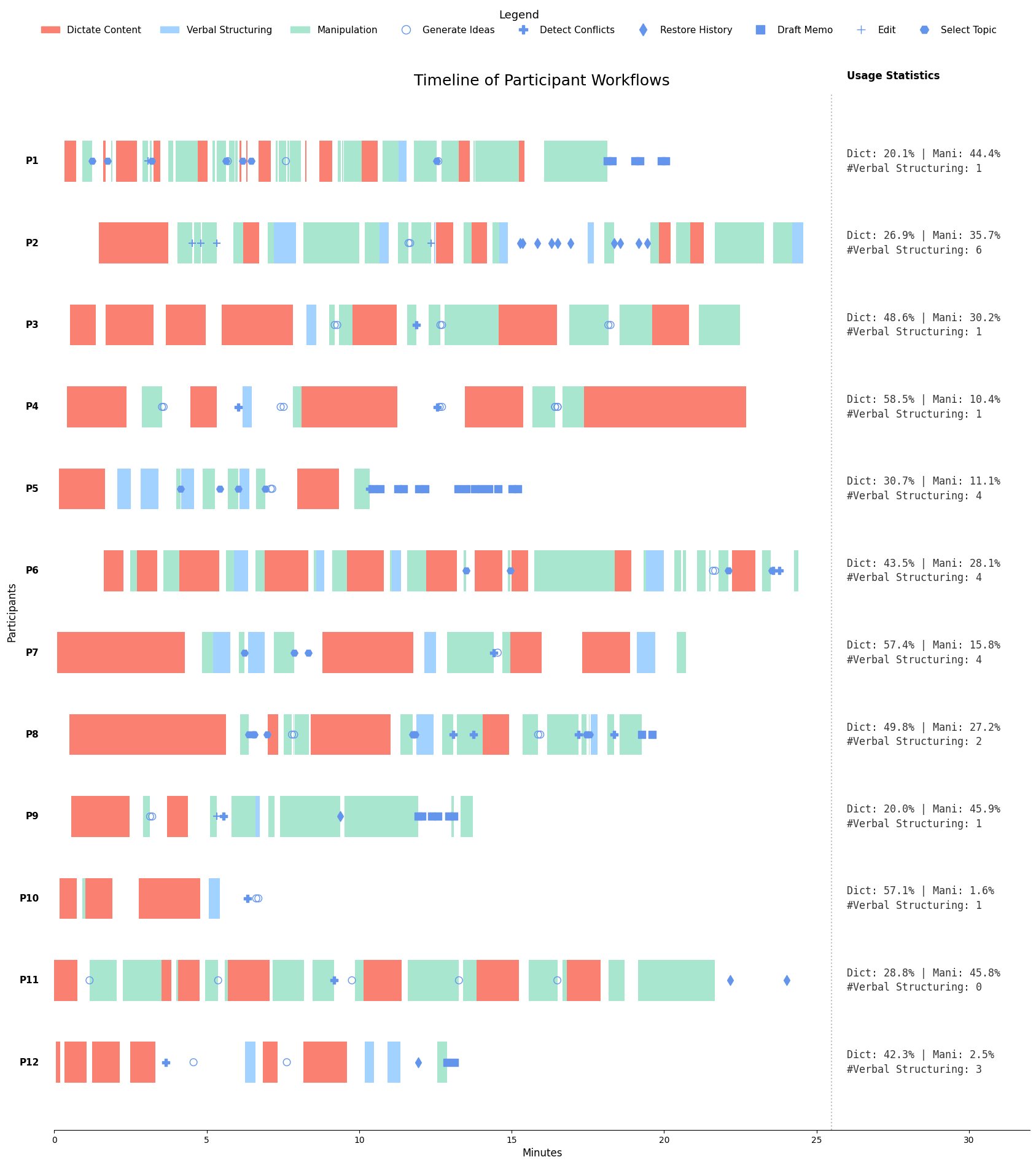}
  \caption{Timelines visualizing participants' workflow for each task. Each use of each feature is annotated as a single point. The speaking and manipulation of nodes is annotated as a time period. \wx{For each participant, the usage statistics are shown on the right, including the time percentage of Dictate Content (Dict) and Manipulation (Mani), and the number of Verbal Structuring (\#Verbal Structuring). The time percentage is calculated as the total time the activity takes up in the entire session.}}
  \Description{A timeline visualization showing the workflow activities of 12 participants (P1-P12) over approximately 25 minutes. Colored segments distinguish primary modes of interaction—Dictate Content (red), Verbal Structuring (blue), and Manipulation (green)—while icons mark discrete events such as Generate Ideas, Detect Conflicts, and Restore History. The right-hand column details Usage Statistics for each participant, specifically the percentage of time allocated to dictation versus manipulation and the total count of verbal structuring commands used.}
  \label{fig:userlog}
\end{figure*}

\twocolumn

\section{Participant Workflows} \label{sec:workflows}
Figure~\ref{fig:userlog} shows a visualization illustrating feature usage of \systemname on timelines of experimental trials. Here we describe the overall workflows and strategies adopted by each participant. 
\begin{itemize}
    \item P1's workflow was highly iterative and tactile, centered on manual curation. They began by speaking in short, focused bursts to add individual skills and career goals to the canvas. After each small addition of content, they spent significant time manually arranging the nodes, using the spatial layout to build connections and structure their thoughts. This pattern of speaking a little, then manually organizing a lot, was their dominant mode of interaction. The AI was only engaged at the very end with a voice command to break their large, manually-created cluster into more granular sub-topics, serving as a final "clean up" step rather than a primary organizational tool.
    \item P2 treated the tool as a narrative assistant, employing a workflow that was almost entirely language-driven with minimal manual manipulation. They began by verbally dictating their academic and professional history. They then used a series of precise text and voice instructions to have the AI organize these facts into clear categories (e.g., ``Research Assistant Experience,'' ``Professional Work Experience''). Critically, they also used the AI to answer their questions, asking it to synthesize their experiences and provide reasons why they were a good fit for the target role. Their use of the ``Thought Evolution'' feature indicates an iterative, conversational approach, trying different commands to find the most effective AI-generated structure and summary.
    \item P3 adopted a problem-solving narrative workflow, relying heavily on speech with little manual node arrangement. They began by verbally outlining a clear problem (a rejected academic paper) and its context. They then used further spoken inputs to progressively deepen their understanding by breaking down the specific reviewer feedback and brainstorming potential solutions. Their primary interaction was talking through the problem space, letting the AI organize the information as it was generated. Voice instructions were used for high-level reorganization to ``make it look clear,'' but only after a significant portion of their thought process had been verbally externalized.
    \item P4 employed a ``funneling'' strategy to achieve focus. They began with a comprehensive "brain dump," verbally externalizing all facets of their current life and career uncertainty. After a brief period of manual arrangement, they used a key voice instruction to direct the AI to prune the canvas, asking it to hide topics related to visa logistics and focus only on their ``post-graduation uncertainty.'' This use of the AI for scoping and managing attention was central to their workflow. Once the canvas was focused on the core problem, they proceeded with a deep and detailed verbal exploration of that topic, relying on the tool to capture their elaborate thoughts without needing further significant manipulation.
    \item P5's workflow was characterized by a dialogue of corrective, hierarchical instructions with almost no manual manipulation. After an initial spoken brain dump, they used a series of precise voice commands to direct the AI in structuring their notes. When the AI’s first attempt at reorganization was not quite right, the user did not resort to manual fixing; instead, they immediately issued a clearer, more specific voice instruction, effectively refining their command to get the desired result. They repeated this pattern at a more granular level, selecting a single topic and issuing another command to restructure just its content. This demonstrates a sophisticated, language-driven approach where the user treats the AI as an organizational assistant that can be guided through iterative verbal commands.
    \item P6 built their argument incrementally, using AI for step-by-step categorization. Unlike users who start with a large brain dump, this user spoke in a series of short, focused bursts, adding one or two related ideas at a time (e.g., a pro and a con). After adding the points for a single line of reasoning, they would issue a simple voice instruction like ``Organize the first reason into a new subtopic'' to have the AI cluster those points together. This additive process of speaking a little, then commanding a small organization was repeated until they had fully constructed their comparative analysis. This workflow also involved lots of manual manipulation.
    \item P7 demonstrated a highly structured, language-driven workflow with almost no manual node manipulation. After an initial, comprehensive "brain dump," they immediately used a series of specific voice instructions to architect the canvas. Their process was notably hierarchical: first, they commanded a broad, top-down reorganization of all content into an analytical framework (e.g., ``Background,'' ``Challenges,'' ``Approaches''). Then, they ``zoomed in,'' selecting a single new section and issuing another granular command to restructure its internal layout. This pattern of using precise voice instructions for both macro and micro-organization shows a strategy that bypasses manual arrangement in favor of verbally directing the AI to build a complex, multi-layered structure.
    \item P8 engaged in a problem-solving narrative, using the tool with minimal reliance on manual arrangement. They began by describing their disconnected projects and a core feeling of confusion. Instead of manually moving nodes to find connections, they used speech to directly ask the canvas for help and leveraged AI-generated questions to guide their reflections. This led to a key conceptual breakthrough. Following this insight, their interaction remained language-based; they issued voice commands for the AI to synthesize their projects under the new theme and to automatically "fix the conflict" in their reasoning. Their workflow was not about visually organizing nodes, but about using conversational and command-based interactions to solve a conceptual problem.
    \item P9 treated the tool primarily as a visual project planning canvas, relying heavily on manual organization while avoiding AI-driven restructuring. They started with a comprehensive ``brain dump'' to externalize all project components. Following this, their dominant activity was the meticulous manual arrangement of nodes to build a visual plan. Notably, when they tested a voice instruction for restructuring, they found the AI's simplified output unsatisfactory and immediately reverted to their preferred, manually curated layout. Their workflow indicates a preference for using speech for initial content capture and manual manipulation for all subsequent organization.
    \item P10 employed a methodical ``speak first, organize later'' workflow with very little manual node manipulation. They began by verbally externalizing all facets of their decision-making problem—listing pros, cons, and lifestyle considerations. Once their thoughts were on the canvas, instead of manually arranging them, they used a single, clear voice instruction (``make the graph clearer'') to have the AI consolidate and structure the entire canvas at once. Their strategy prioritized comprehensive verbal brainstorming followed by a single, decisive AI-powered reorganization.
    \item P11 adopted a balanced, iterative multi-modal strategy. They began with a broad spoken statement and then entered a consistent loop: they would speak to add a new layer of detail, and immediately follow up with significant manual manipulation, dragging and rearranging the new nodes to visually integrate them. This tight cycle of speaking and then manually organizing was their primary method for building out their thoughts. For major restructuring, they used high-level commands to have the AI refactor the entire canvas into a more thematic summary.
    \item P12 engaged in a highly reflective, dialectical process that was almost entirely language-driven, with minimal manual manipulation of nodes. After externalizing their initial thoughts, they relied on a series of progressively more sophisticated voice instructions to direct the AI. They commanded it to not only reorganize their ideas but also to perform analysis by categorizing their thoughts into points of ``overlap'' and ``conflict.'' This demonstrates a workflow where the user treats the AI as an analytical partner, using spoken language for both content generation and complex structural transformations, largely bypassing manual arrangement.
\end{itemize}



\section{Prompts}
\subsection{The default prompt for Baseline}
\label{Baseline default promptl}

At the beginning of the thought organization task with using the Baseline, the following prompt would be sent to ChatGPT, in order to make it comparable with \systemname{}.

\textit{I want to organize my thoughts gradually through some graphics and text information. I will enter my thoughts mainly by speaking. I want you to simulate a tool that can perform the following functions:
1. You can provide a multi-level structure to represent the content of what I just input. The structure can either be a mindmap, an outline, or both. The mindmap should not be similar to the outline; you should use text and dashed lines to simulate a graphic mind map.
2. Every time I enter something new, you should add the new content to the existing content so the whole content is "evolving".}

\textit{When you are ready, I will start.}
\subsection{Prompts Used in \texorpdfstring{\systemname}{Orality}} \label{sec:prompts}
\begin{table*}[ht!]
\centering
\renewcommand{\arraystretch}{1.5}
\begin{tabular}{p{2cm}|p{12cm}}
\hline
\textbf{Tasks} & \textbf{Prompt Template} \\ \hline
Extract Topic and Content Nodes  & You are a listener of someone talking. Analyze the given text and extract the key information in the form of short sentence entities instead of keywords or short phrases. If the subject is the speaker itself, there is no need to add a subject. Some sentence entities could be grouped into the same "point of view", which means the speaker's statement, judgement, opinion, or concern, etc, on something. If the user has some instructions, follow the instructions to create the ``points of view''. Generate one to three sentence entities under each "point of view" so that each independent entity contains complete information and prevents two entities from containing similar information. Don't generate entities that have similar content.
Put them in the list in the following format:
[{
    "topic": ...,
    "entities": [..., ..., ...]
},...] \\ \hline
Reorganization Instructions & You are a content organizer. Analyze the given outline and reorganize the elements in it according to the user's specific instructions. In the current outline, there are some topics and some entities that belong to them. You need to generate a new outline with topics that follow the instructions and pick the existing entities to put them under the new outline, or extract relevant but not existing entities from the user's full text. There are three potential common cases of the instructions:

1. The user asks for a new structure with specific topics, then you should use the user-defined topics as the topics in the output.

2. The user asks for a new structure without specifying new topics, then you should come up with topics relevant to that structure.

3. The user asks for adding new topics based on the current structure, then you should add the new topics to the existing outline and extract new relevant content entities from the full text.

If the user's inquiry is not in these cases, you should treat them as carefully as you can and keep the existing structure as far as you can. Output the new outline in the following format:
[{
    "topic": ...,
    "entities": [
        {
            "type": If this is an existing entity, don't change its type; otherwise, set to 1
            "text": ...
        }
        ...
    ]
},...] \\ \hline
Ask Me Questions & You are a thoughtful conversation facilitator helping users organize their thoughts. Given a topic and its current content, generate 1-2 directing questions that would help the user explore this topic more deeply or provide missing information.

The questions should:

1. Be open-ended and thought-provoking

2. Help the user expand on the topic with specific details, examples, or insights

3. Be directly relevant to the existing content in the topic

4. Encourage the user to share experiences, concrete examples, or deeper analysis

5. Be phrased as helpful prompts (e.g., "What specific challenges...", "Can you describe...", "How did you...")

6. Avoid yes/no questions

7. Be concise but specific

Format the response as a simple JSON array of question strings: ["Question 1?", "Question 2?"]

Generate 1-2 questions per topic (2 if the topic has very little content, 1 if it has some content). \\ \hline
\end{tabular}
\label{tab:prompts}
\end{table*}

\begin{table*}[ht!]
\centering
\renewcommand{\arraystretch}{1.5}
\begin{tabular}{p{2cm}|p{12cm}}
\hline
\textbf{Tasks} & \textbf{Prompt Template} \\ \hline
Find Conflicts & You are an expert at identifying logical conflicts and contradictions in thoughts and ideas. 

Analyze pairs of content for conflicts such as:

1. Direct contradictions - statements that directly oppose each other

2. Logical inconsistencies - statements that cannot both be true

3. Value conflicts - opposing values, priorities, or principles

4. Strategy conflicts - contradictory approaches to the same problem

5. Assumption conflicts - incompatible underlying assumptions

For each pair, determine:

- Is there a meaningful conflict? (not just different topics or perspectives)

- What type of conflict is it?

- How confident are you? (1-10 scale)

- Brief explanation of why it's a conflict

Respond with ONLY a JSON array (no markdown formatting) where each element represents a detected conflict:

[
    {
        "pair\_id": 0,
        "has\_conflict": true,
        "conflict\_type": "Direct Contradiction",
        "confidence": 8,
        "reason": "One states X is true while the other states X is false"
    }
]

Only include pairs where you detect a genuine conflict (has\_conflict: true). Be selective - different perspectives on the same topic are not necessarily conflicts. Focus on substantive contradictions that would create logical problems if both were true. Return an empty array [] if no conflicts are found.
\\ \hline

Export Reports & Please create a professional memo based on the following organized content and the selected style:
'comprehensive': You are helping a user organize and synthesize their own thoughts.  Create a comprehensive personal summary that:

1. Synthesizes the key themes and insights from their content

2. Identifies important patterns and connections in their thinking

3. Highlights significant insights and potential next steps

4. Uses a personal, reflective tone (address the user directly)

5. Organizes thoughts into clear, logical sections

Structure as: Key Themes, Important Insights, Connections \& Patterns, and Next Steps.

'executive': Create a high-level personal summary of the user's thoughts. Focus on:

1. The most important themes and insights from their content

2. Key decisions or actions they might want to consider

3. Strategic implications of their thinking

4. Personal, direct language (use 'you' and 'your')

5. Concise but comprehensive overview

Keep it brief but capture the essential insights from their thinking.

'bullet': Create a structured personal summary using clear bullet points:

1. Organize the user's thoughts into logical groups

2. Use clear, scannable bullet points

3. Include action items or next steps where relevant

4. Use personal language (address the user directly)

5. Make it easy to scan and reference later

Create a cohesive, well-structured memo that synthesizes these topics and provides valuable insights. The memo should be thorough but readable, highlighting key themes, relationships between topics, and actionable insights.
\\ \hline
\end{tabular}
\label{tab:prompts2}
\end{table*}


\end{document}